\documentclass[journal]{IEEEtran}
\usepackage{cite}
\usepackage{amsmath,amssymb,amsfonts}
\usepackage{algorithm}
\usepackage{algpseudocode}
\usepackage{graphicx}
\usepackage{textcomp}
\usepackage{xcolor}
\usepackage{enumitem}
\usepackage{tikz}
\usepackage{url}
\usetikzlibrary{arrows.meta}
\usepackage{adjustbox}
\usepackage{orcidlink}
\usepackage[acronym,toc]{glossaries}
\usepackage{hyperref}
\usepackage{color,soul}
\soulregister\gls{7}
\soulregister\glspl{7}
\usepackage[caption=false,font=normalsize,labelfont=sf,textfont=sf]{subfig}
\usepackage[acronym,nonumberlist]{glossaries-extra}
\usepackage{multicol,multirow}
\usepackage{booktabs}
\makeglossaries
\setabbreviationstyle[acronym]{long-short}

\newacronym{2d}{2D}{two-dimensional}
\newacronym{3d}{3D}{three-dimensional}
\newacronym{3gpp}{3GPP}{3rd generation partnership project}
\newacronym{3msk}{3MSK}{3-level minimum-shift-keying}
\newacronym{4g}{4G}{fourth-generation}
\newacronym{5g}{5G}{fifth-generation}
\newacronym{6dof}{6DoF}{six degrees of freedom}
\newacronym{6g}{6G}{sixth-generation}
\newacronym{6tisch}{6TiSCH}{IPv6 over the TSCH mode of IEEE 802.15.4e}
\newacronym{abp}{ABP}{authentication by personalisation}
\newacronym{ac}{AC}{alternating current}
\newacronym{aci}{ACI}{adjacent channel interference}
\newacronym{aclr}{ACLR}{adjacent channel leakage ratio}
\newacronym{acpr}{ACPR}{adjacent channel power ratio}
\newacronym{adc}{ADC}{analog-to-digital converter}
\newacronym{adr}{ADR}{adaptive data rate}
\newacronym{aec}{AEC}{aluminum electrolytic capacitor}
\newacronym{aes}{AES}{Advanced Encryption Standard}
\newacronym{afe}{AFE}{analog front end}
\newacronym{ag}{AG}{array gain}
\newacronym{agc}{AGC}{automatic gain controller}
\newacronym{agps}{A-GPS}{Assisted Global Positioning System} 
\newacronym{ai}{AI}{artificial intelligence}
\newacronym{aimd}{AIMD}{Additive-Increase/Multiplicative-Decrease}
\newacronym{aiot}{A-IoT}{Ambient IoT}
\newacronym{alk}{ALK}{Alkaline}
\newacronym{am}{AM}{amplitude modulation}
\newacronym{amam}{AM/AM}{amplitude modulation to amplitude modulation}
\newacronym{ambc}{AmBC}{ambient backscatter communication}
\newacronym{amc}{AMC}{automatic modulation classification}
\newacronym{amp}{AMP}{AMbient Power}
\newacronym{ampm}{AM/PM}{amplitude modulation to phase modulation}
\newacronym{aoa}{AOA}{angle-of-arrival}
\newacronym{aod}{AOD}{angle-of-departure}
\newacronym{ap}{AP}{access point}
\newacronym{api}{API}{application program interface}
\newacronym{apt}{APT}{acoustic power transfer}
\newacronym{apu}{APU}{access point unit}
\newacronym{ar}{AR}{augmented reality}
\newacronym{arima}{ARIMA}{Auto-Regressive Integrated Moving Average}
\newacronym{arp}{ARP}{Antenna Reference Point}
\newacronym{asic}{ASIC}{application specific integrated circuit}
\newacronym{ask}{ASK}{amplitude-shift keying}
\newacronym{at}{AT}{ATtention}
\newacronym{auv}{AUV}{autonomous underwater vehicle}
\newacronym{awg}{AWG}{arbitrary waveform generator}
\newacronym{awgn}{AWGN}{additive white Gaussian noise}
\newacronym{baw}{BAW}{bulk acoustic wave}
\newacronym{bb}{BB}{base-band}
\newacronym{bc}{BC}{backscatter communication}
\newacronym{bcjr}{BCJR}{Bahl-Cocke-Jelinek-Raviv}
\newacronym{bd}{BD}{backscatter device}
\newacronym{be}{BE}{Belgium}
\newacronym{ber}{BER}{bit error rate}
\newacronym{bf}{BF}{beamforming}
\newacronym{bga}{BGA}{ball grid array}
\newacronym{bjt}{BJT}{bipolar junction transistor}
\newacronym{bldc}{BLDC}{brushless DC}
\newacronym{ble}{BLE}{Bluetooth low energy}
\newacronym{bler}{BLER}{block error rate}
\newacronym{bod}{BOD}{Brown-Out Detection}
\newacronym{bom}{BOM}{bill of materials}
\newacronym{bpsk}{BPSK}{binary phase-shift keying}
\newacronym{brp}{BRP}{beam refinement process}
\newacronym{bs}{BS}{base station}
\newacronym{btwt}{bTWT}{broadcast TWT}
\newacronym{bu}{BU}{booster unit}
\newacronym{bw}{BW}{bandwidth}
\newacronym{ca}{CA}{carrier emitter}
\newacronym{cad}{CAD}{channel activity detection}
\newacronym{cars}{CARS}{calibration reference signal}
\newacronym{cbm}{CBM}{condition based maintenance}
\newacronym{cbra}{CBRA}{contention-based random access}
\newacronym{cc}{CC}{constant current}
\newacronym{ccdf}{CCDF}{complementary cumulative distribution function}
\newacronym{ccnn}{CCNN}{circular convolutional neural network}
\newacronym{ccs}{CCS}{correlative channel sounder}
\newacronym{cdf}{CDF}{cumulative distribution function}
\newacronym{cdma}{CDMA}{code division-multiple access}
\newacronym{cdrx}{CDRX}{connected mode DRX}
\newacronym{ce}{CE}{coverage enhancement}
\newacronym{ced}{CED}{cumulative energy density}
\newacronym{ceofdm}{CE-OFDM}{constant-envelope OFDM}
\newacronym{cept}{CEPT}{European Conference of Postal and Telecommunications Administrations}
\newacronym{cf}{CF}{cell-free}
\newacronym{cfmmimo}{CF-mMIMO}{cell-free massive MIMO}
\newacronym{cfo}{CFO}{carrier frequency offset}
\newacronym{cfra}{CFRA}{contention-free random access}
\newacronym{cir}{CIR}{channel impulse response}
\newacronym{cla}{CLA}{closed-loop approach}
\newacronym{clk}{CLK}{clock}
\newacronym{cmos}{CMOS}{complementary metal oxide semiconductor}
\newacronym{cn}{CN}{Core network}
\newacronym{cnn}{CNN}{convolutional neural network}
\newacronym{co}{CO}{continuous operation}
\newacronym{cordic}{CORDIC}{coordinate rotation digital computer}
\newacronym{cots}{COTS}{commercial off-the-shelf}
\newacronym{cp}{CP}{cyclic prefix}
\newacronym{cpe}{CPE}{common phase error}
\newacronym{cpfsk}{CPFSK}{continuous phase frequency shift keying}
\newacronym{cpm}{CPM}{continuous phase modulation}
\newacronym{cpo}{CPO}{carrier phase offset}
\newacronym{cpt}{CPT}{capacitive power transfer}
\newacronym{cpw}{CPW}{coplanar waveguide}
\newacronym{cqi}{CQI}{channel quality indicator}
\newacronym{cr}{CR}{coding rate}
\newacronym{crc}{CRC}{cyclic redundancy check}
\newacronym{crlb}{CRLB}{Cram\'er-Rao lower bound}
\newacronym{crs}{CRS}{cell reference signal}
\newacronym{crtwt}{C-RTWT}{coordinated restricted TWT}
\newacronym{cs}{CS}{compressed sensing}
\newacronym{cse}{CSE}{channel state estimation}
\newacronym{csi}{CSI}{channel state information}
\newacronym{csp}{CSP}{contact service point}
\newacronym{css}{CSS}{chirp spread spectrum}
\newacronym{cu}{CU}{central unit}
\newacronym{cv}{CV}{constant voltage}
\newacronym{cw}{CW}{continuous wave}
\newacronym{d2d}{D2D}{device-to-device}
\newacronym{d2r}{D2R}{device-to-reader}
\newacronym{dab}{DAB}{Digital Audio Broadcasting}
\newacronym{dac}{DAC}{digital-to-analog converter}
\newacronym{daq}{DAQ}{data acquisition system}
\newacronym{das}{DAS}{distributed antenna systems}
\newacronym{db}{DB}{duty-cycled barebone}
\newacronym{dbpsk}{DBPSK}{Differential Binary Phase Shift Keying}
\newacronym{dc}{DC}{direct current}
\newacronym{dcc}{DCC}{dynamic cooperation clustering}
\newacronym{ddc}{DDC}{digital down conversion}
\newacronym{de}{DE}{drain efficiency}
\newacronym{dect}{DECT}{Digital Cordless Telecommunications}
\newacronym{dft}{DFT}{discrete Fourier transform}
\newacronym{dftsofdm}{DFT-s-OFDM}{discrete Fourier Transform spread OFDM}
\newacronym{dftsofdmfdss}{DFT-s-OFDM-FDSS}{DFT-s-OFDM with frequency-domain spectral shaping}
\newacronym{dftsofdmfdssse}{DFT-s-OFDM-FDSS-SE}{DFT-s-OFDM with FDSS and spectral extension}
\newacronym{dl}{DL}{downlink}
\newacronym{dlc}{DLC}{Distributed Laser Charging}
\newacronym{dli}{DLI}{direct link interference}
\newacronym{dlt}{DLT}{Distributed Ledger Technology}
\newacronym{dm}{DM}{diffuse multipath}
\newacronym{dma}{DMA}{Direct Memory Access}
\newacronym{dmac}{DMAC}{Direct Memory Access Controller}
\newacronym{dmc}{DMC}{diffuse multipath component}
\newacronym{dmimo}{D-MIMO}{distributed MIMO}
\newacronym{dnn}{DNN}{deep neural network}
\newacronym{doa}{DOA}{direction-of-arrival}
\newacronym{dp}{DP}{duty-cycled persistent}
\newacronym{dpd}{DPD}{digital pre-distortion}
\newacronym{dpdk}{DPDK}{Data Plane Development Kit}
\newacronym{dram}{DRAM}{dynamic random-access memory}
\newacronym{drcs}{$\Delta$RCS}{differential-radar cross section}
\newacronym{drx}{DRX}{Discontinuous Reception Mode}
\newacronym{dsb}{DSB}{double-sideband}
\newacronym{dsl}{DSL}{digital subscriber line}
\newacronym{dsp}{DSP}{digital signal processing}
\newacronym{dss}{DSS}{Dataset Storage Standard}
\newacronym{duc}{DUC}{digital up-converter}
\newacronym{dvb}{DVB}{Digital Video Broadcasting}
\newacronym{de-ewma}{DE-EWMA}{Dynamic Error-Exponentially Weighted Moving Average}
\newacronym{e2e}{E2E}{end-to-end}
\newacronym{easa}{EASA}{European Union Aviation Safety Agency}
\newacronym{ebg}{EBG}{electromagnetic bandgap}
\newacronym{ebw}{EBW}{excess bandwidth}
\newacronym{ec}{EC}{European Commission}
\newacronym{ecc}{ECC}{Electronic Communications Committee}
\newacronym{ecdf}{eCDF}{empirical cumulative distribution function}
\newacronym{ecdlp}{ECDLP}{elliptic curve discrete logarithm problem}
\newacronym{ecsp}{ECSP}{edge computing service point}
\newacronym{edlc}{EDLC}{electrostatic double-layer capacitors}
\newacronym{edrx}{eDRX}{Extended Discontinuous Reception Mode}
\newacronym{ee}{EE}{energy efficiency}
\newacronym{egprs}{EGPRS}{Enhanced Data Rates for GSM Evolution}
\newacronym{eh}{EH}{energy harvesting}
\newacronym{eirp}{EIRP}{equivalent isotropically radiated power}
\newacronym{em}{EM}{electromagnetic}
\newacronym{embb}{eMBB}{enhanced Mobile Broadband}
\newacronym{en}{EN}{energy neutral}
\newacronym{end}{END}{energy-neutral device}
\newacronym{enob}{ENOB}{effective number of bits}
\newacronym{eol}{EoL}{end of life}
\newacronym{ep}{EP}{energy profiler}
\newacronym{epd}{EPD}{electronic paper display}
\newacronym{epu}{EPU}{edge processing unit}
\newacronym{er}{ER}{Energy Receiver}
\newacronym{erc}{ERC}{European Radiocommunications Committee}
\newacronym{erp}{ERP}{effective radiation power}
\newacronym[plural=ESCs,firstplural=electronic speed controllers (ESCs)]{esc}{ESC}{electronic speed control}
\newacronym{esd}{ESD}{electrostatic discharge}
\newacronym{esl}{ESL}{electronic shelf label}
\newacronym{esr}{ESR}{equivalent series resistance}
\newacronym{et}{ET}{Energy Transmitter}
\newacronym{etsi}{ETSI}{European Telecommunications Standards Institute}
\newacronym{evd}{EVD}{eigenvalue decomposition}
\newacronym{evm}{EVM}{Error Vector Magnitude}
\newacronym{ewlb}{eWLB}{Embedded Wafer Level Ball Grid Array}
\newacronym{ewma}{EWMA}{Exponentially Weighted Moving Average}
\newacronym{eipm}{EIPM}{Energy Intelligence Platform Module}
\newacronym{fa}{FA}{federation anchor}
\newacronym{fair}{FAIR}{Findability, Accessibility, Interoperability, and Reuse of digital assets}
\newacronym{fc}{FC}{fusion center}
\newacronym{fcc}{FCC}{Federal Communications Commission}
\newacronym{fcf}{FCF}{frequency correlation function}
\newacronym{fd}{FD}{front-haul distance}
\newacronym{fdd}{FDD}{frequency-division duplexing}
\newacronym{fde}{FDE}{frequency domain equalizer}
\newacronym{fdm}{FDM}{frequency-division multiplexing}
\newacronym{fdma}{FDMA}{frequency division multiple access}
\newacronym{fdss}{FDSS}{frequency-domain spectral shaping}
\newacronym{fem}{FEM}{finite element analysis}
\newacronym{fembb}{feMBB}{further enhanced mobile broadband}
\newacronym{fft}{FFT}{fast Fourier transform}
\newacronym{fh}{FH}{fronthaul}
\newacronym{fhss}{FHSS}{frequency hopping spread spectrum}
\newacronym{fifo}{FIFO}{First In, First Out}
\newacronym{fim}{FIM}{Fisher information matrix}
\newacronym{fits}{FITS}{Flexible Image Transport System}
\newacronym{fl}{FL}{Federated learning}
\newacronym{fom}{FoM}{figure of merit}
\newacronym{fov}{FoV}{field of view}
\newacronym{fpga}{FPGA}{field-programmable gate array}
\newacronym{fqt}{FQT}{Fully Quantized Training}
\newacronym{fr1}{FR1}{frequency range 1}
\newacronym{fr2}{FR2}{frequency range 2}
\newacronym{fram}{FRAM}{Ferroelectric Random Access Memory}
\newacronym{fsk}{FSK}{frequency shift keying}
\newacronym{fspl}{FSPL}{free space path loss}
\newacronym{fss}{FSS}{frequency selective surface}
\newacronym{gan}{GaN}{gallium nitride}
\newacronym{gb}{GB}{grant-based}
\newacronym{gdpr}{GDPR}{general data protection regulation}
\newacronym{gf}{GF}{grant-free}
\newacronym{gmsk}{GMSK}{Gaussian minimum-shift keying}
\newacronym{gnb}{gNB}{Next Generation Node B}
\newacronym{gni}{GNI}{gross national income}
\newacronym{gnn}{GNN}{graph neural network}
\newacronym{gnss}{GNSS}{global navigation satellite system}
\newacronym{gpclk}{GPCLK}{general purpose clock}
\newacronym{gpio}{GPIO}{General-Purpose Input/Output}
\newacronym{gpl}{GPL}{GNU General Public License}
\newacronym{gprs}{GPRS}{General Packet Radio Services}
\newacronym{gps}{GPS}{Global Positioning System}
\newacronym{gpu}{GPU}{graphical processing unit}
\newacronym{grc}{GRC}{GNU Radio Companion}
\newacronym{gscm}{GSCM}{geometry‐based stochastic model}
\newacronym{gsm}{GSM}{Global System for Mobile Communications}  %
\newacronym{gspm}{GSpM}{Generalized spatial modulation}
\newacronym{gwp}{GWP}{Global Warming Potential}
\newacronym{har}{HAR}{human activity recognition}
\newacronym{harq}{HARQ}{hybrid automatic repeat request}
\newacronym{hat}{HAT}{hardware attached on top}
\newacronym{hcs}{HCS}{human-centric services}
\newacronym{hdf5}{HDF5}{Hierarchical Data Format version 5}
\newacronym{hdr}{HDR}{high data rate}
\newacronym{hfss}{HFSS}{High Frequency Simulator Software}
\newacronym{hf}{HF}{high frequency}
\newacronym{hmd}{HMD}{head-mounted display}
\newacronym{hpbm}{HPBM}{half power beam width}
\newacronym{HyMPRo}{HyMPRo}{Hybrid Multi-Path Routing algorithm}
\newacronym{i.i.d.}{i.i.d.}{independent and identically distributed}
\newacronym{i2c}{I2C}{Inter-Integrated Circuit}
\newacronym{iaq}{IAQ}{Indoor Air Quality}
\newacronym{ib}{IB}{in-band}
\newacronym{ibbc}{IBBC}{inter-band beam configuration}
\newacronym{ibo}{IBO}{input back-off}
\newacronym{ic}{IC}{integrated circuit}
\newacronym{iccs}{ICCS}{Ilmsens correlative channel sounder}
\newacronym{ici}{ICI}{intercarrier interference}
\newacronym{icnirp}{ICNIRP}{International Commission on Non-Ionizing Radiation Protection}
\newacronym{id}{ID}{information decoding}
\newacronym{idft}{IDFT}{inverse discrete Fourier transform}
\newacronym{idxm}{IDXM}{index modulation}
\newacronym{ieee}{IEEE}{Institute of Electrical and Electronics Engineers}
\newacronym{if}{IF}{intermediate-frequency}
\newacronym{ifft}{IFFT}{inverse fast-Fourier-transform}
\newacronym{iid}{i.i.d.}{independently and identically distributed}
\newacronym{iiot}{IIoT}{Industrial IoT}
\newacronym{iis}{IIS}{integrated information system}
\newacronym{im}{IM}{intermodulation}
\newacronym{imd}{IMD}{intermodulation distortion}
\newacronym{imu}{IMU}{inertial measurement unit}
\newacronym{inh}{InH}{indoor hotspot office}
\newacronym{io}{IO}{input/output}
\newacronym{ioe}{IoE}{Internet of Everything}
\newacronym{iot}{IoT}{Internet of Things}
\newacronym{ipt}{IPT}{inductive power transfer}
\newacronym{ipy}{IPY}{Interventions per Year}
\newacronym{iq}{IQ}{in-phase and quadrature}
\newacronym{iqi}{IQI}{IQ imbalance}
\newacronym{ir}{IR}{infrared}
\newacronym{isac}{ISAC}{integrated sensing and communication} 
\newacronym{isi}{ISI}{intersymbol interference}
\newacronym{ism}{ISM}{industrial, scientific and medical}
\newacronym{isp}{ISP}{internet service provider}
\newacronym{itu}{ITU}{International Telecommunication Union}
\newacronym{itu-r}{ITU-R}{ITU – Radiocommunication Sector}
\newacronym{jesd}{JESD}{Joint Electron Devices Engineering Council}
\newacronym{jfet}{JFET}{junction field effect transistor}
\newacronym{kpi}{KPI}{key performance indicator}
\newacronym{ktofdm}{KT-DFT-s-OFDM}{known-tail-DFT-s-OFDM}
\newacronym{kvi}{KVI}{key value indicator}
\newacronym{larva}{LARVA}{LARge Virtual Array}
\newacronym{lbt}{LBT}{Listen-Before-Talk}
\newacronym{lca}{LCA}{life cycle assessment}
\newacronym{lch}{LCH}{logical channel}
\newacronym{lco}{LCO}{lithium cobalt oxide}
\newacronym{ldo}{LDO}{low-dropout voltage regulator}
\newacronym{ldpc}{LDPC}{low-density parity-check}
\newacronym{ldr}{LDR}{low data rate}
\newacronym{led}{LED}{Light Emitting Diode}
\newacronym{less}{LESS}{Low Energy Scheduler Solution}
\newacronym{lev}{LEV}{light electric vehicle}
\newacronym{lfp}{LFP}{lithium iron phosphate}
\newacronym{lib}{LIB}{Lithium-Ion Battery}
\newacronym{lic}{LIC}{lithium-ion capacitor}
\newacronym{lid}{LID}{Lithium Iron Disulfide}
\newacronym{lidar}{LiDAR}{light detection and ranging}
\newacronym{liion}{Li-ion}{lithium-ion}
\newacronym{lipo}{LiPo}{lithium polymer}
\newacronym{lis}{LIS}{large intelligent surface}
\newacronym{llh}{LLH}{log-likelihood}
\newacronym{lls}{LLS}{link-level simulation}
\newacronym{lmd}{LMD}{Lithium Manganese Dioxide}
\newacronym{lmmse}{LMMSE}{least minimum mean square error}
\newacronym{lmo}{LMO}{lithium ion manganese oxide}
\newacronym{lna}{LNA}{low-noise amplifier}
\newacronym{lo}{LO}{local oscillator}
\newacronym{lora}{LoRa}{long range}
\newacronym{lorawan}{LoRaWAN}{long-range wide-area network}
\newacronym{los}{LoS}{line-of-sight}
\newacronym{lp}{LP}{linear programming}
\newacronym{lpf}{LPF}{low-pass filter}
\newacronym{lpt}{LPT}{laser power transfer}
\newacronym{lpwa}{LPWA}{Low Power Wide Area}
\newacronym{lpwan}{LPWAN}{low-power wide-area network}
\newacronym{lpwans}{LPWANs}{Low-Power Wide-Area Networks}
\newacronym{lqi}{LQI}{link quality indicator}
\newacronym{llr}{LLR}{log-likelihood ratio}
\newacronym{lrelu}{LReLU}{leaky rectified linear unit}
\newacronym{lrt}{LRT}{likelihood-ratio test}
\newacronym{ls}{LS}{least squares}
\newacronym{lsa}{LSA}{large synthetic array}
\newacronym{lsf}{LSF}{large-scale fading}
\newacronym{lsfc}{LSFC}{large-scale fading component}
\newacronym{lstm}{LSTM}{Long Short-Term Memory}
\newacronym{ltc}{LTC}{lithium thionyl chloride}
\newacronym{lte}{LTE}{Long Term Evolution}
\newacronym{ltem}{LTE-M}{Long-Term Evolution Machine Type Communication}
\newacronym{lti}{LTI}{linear time-invariant}
\newacronym{lto}{LTO}{lithium titanate}
\newacronym{lusta}{LUSTA 5G }{Logistique mUltimodale Sécuritaire Téléopérée \& Autonome 5G}
\newacronym{m2m}{M2M}{machine to machine}
\newacronym{ma}{MA}{Movable Antennas}
\newacronym{mac}{MAC}{Medium Access Control}
\newacronym{mate}{MATE}{millimeter-wave MIMO testbed}
\newacronym{mc}{MC}{Monte Carlo}
\newacronym{mcl}{MCL}{Maximum Coupling Loss}
\newacronym{mcs}{MCS}{modulation and coding scheme}
\newacronym{mcu}{MCU}{microcontroller unit}
\newacronym{mec}{MEC}{multi-access edge computing}
\newacronym{mems}{MEMS}{micro-electromechanical systems}
\newacronym{mf}{MF}{matched filter}
\newacronym{milp}{MILP}{Mixed Integer Linear Programming}
\newacronym{mimo}{MIMO}{multiple-input multiple-output}
\newacronym{miso}{MISO}{multiple-input single-output}
\newacronym{ml}{ML}{machine learning}
\newacronym{mlp}{MLP}{multilayer perceptron}
\newacronym{mmic}{MMIC}{monolithic microwave integrated circuit}
\newacronym{mmimo}{mMIMO}{massive MIMO}
\newacronym{mmse}{MMSE}{minimum mean square error}
\newacronym{mmtc}{mMTC}{massive machine-type communications}
\newacronym{mmwave}{mmWave}{millimeter wave}
\newacronym{mn}{MN}{matching network}
\newacronym{mosfet}{MOSFET}{metal-oxide semiconductor field effect transistor}
\newacronym{mpc}{MPC}{multipath component}
\newacronym{mpp}{MPP}{magnetic power profile}
\newacronym{mppt}{MPPT}{maximum power point tracking}
\newacronym{mqtt}{MQTT}{Message Queuing Telemetry Transport}
\newacronym{mr}{MR}{maximum ratio}
\newacronym{mrc}{MRC}{maximum ratio combining}
\newacronym{mrc_em}{MRC}{maximum ratio combining}
\newacronym{mrc_EM}{MRC}{Magnetic Resonance Coupling}
\newacronym{mrt}{MRT}{maximum ratio transmission}
\newacronym{mse}{MSE}{mean square error}
\newacronym{msk}{MSK}{Minimum-Shift Keying}
\newacronym{mtc}{MTC}{Machine-Type Communication}
\newacronym{multi-rat}{Multi-RAT}{multiple radio access technology}
\newacronym{multirat}{Multi-RAT}{Multiple Radio Access Technology}
\newacronym{music}{MUSIC}{MUltiple SIgnal Classification}
\newacronym{nas}{NAS}{Neural Architecture Search}
\newacronym{navauwall}{NAVAUWALL}{AUtomated NAVigation in WALLonia}
\newacronym{nb}{NB}{narrowband}
\newacronym{nbiot}{NB-IoT}{narrowband IoT}
\newacronym{nca}{NCA}{nickel cobalt aluminum}
\newacronym{netcdf}{NetCDF}{Network Common Data Form}
\newacronym{nf}{NF}{noise figure}
\newacronym{nfc}{NFC}{near-field communication}
\newacronym{nfv}{NFV}{network function virtualization}
\newacronym{ngmn}{NGMN}{Next Generation Mobile Networks }
\newacronym{ni}{NI}{National Instruments}
\newacronym{nicd}{NiCd}{nikkel cadmium}
\newacronym{nimh}{NiMH}{nikkel metal hydride}
\newacronym{nlos}{NLoS}{non-line-of-sight}
\newacronym{nmc}{NMC}{nickel manganese cobalt}
\newacronym{nmos}{nMOS}{n-channel metal-oxide semiconductor}
\newacronym{nn}{NN}{neural network}
\newacronym{nnls}{NNLS}{non-negative least squares}
\newacronym{noma}{NOMA}{non-orthogonal multiple access}
\newacronym{np}{NP}{Neyman-Pearson}
\newacronym{npbch}{NPBCH}{Narrowband Physical Broadcast Channel}
\newacronym{npss}{NPSS}{Narrow Band Primay Synchronization Signal}
\newacronym{nr}{NR}{New Radio}
\newacronym{nrs}{NRS}{Narrow Band Reference Signal}
\newacronym{nsss}{NSSS}{Narrowband Secondary Synchronization Signal}
\newacronym{ntp}{NTP}{network time protocol}
\newacronym{oai}{OAI}{OpenAirInterface} 
\newacronym{obw}{OBW}{occupied bandwidth}
\newacronym{odi}{O-DI}{On-Device Intelligence}
\newacronym{oef}{OEF}{Organisation Environmental Footprint}
\newacronym{ofa}{OFA}{Once-for-All}
\newacronym{ofdm}{OFDM}{orthogonal frequency-division multiplexing}
\newacronym{ofdma}{OFDMA}{orthogonal frequency-division multiple access}
\newacronym{ofdmim}{OFDM-IM}{OFDM with index modulation}
\newacronym{olos}{OLoS}{obstructed-line-of-sight}
\newacronym{oma}{OMA}{orthogonal multiple access}
\newacronym{oob}{OOB}{out-of-band}
\newacronym{ook}{OOK}{on-off keying}
\newacronym{opbo}{OPBO}{output power backoff}
\newacronym{oran}{O-RAN}{open radio-access network}
\newacronym{os}{OS}{operating system}
\newacronym{ota}{OTA}{over-the-air}
\newacronym{otaa}{OTAA}{over-the-air authentication}
\newacronym{otaml}{OTA-TinyML}{Over-the-air TinyML}
\newacronym{ova}{OVA}{One-Versus-All}
\newacronym{p1}{P1}{Phase 1}
\newacronym{p2}{P2}{Phase 2}
\newacronym{p2p}{P2P}{point-to-point}
\newacronym{pa}{PA}{power amplifier}
\newacronym{pae}{PAE}{power-added efficiency}
\newacronym{pam}{PAM}{pulse amplitude modulation}
\newacronym{pana}{PanA}{Panel A}
\newacronym{panb}{PanB}{Panel B}
\newacronym{papr}{PAPR}{peak-to-average power ratio}
\newacronym{pb}{PB}{power beacon}
\newacronym{pc}{PC}{pilot count}
\newacronym{pcb}{PCB}{printed circuit board}
\newacronym{pcg}{PCG}{power consumption gain}
\newacronym{pcie}{PCIe}{Peripheral Component Interconnect Express}
\newacronym{pcsi}{PCSI}{perfect channel state information}
\newacronym{pd}{PD}{powered device}
\newacronym{pdcch}{PDCCH}{physical downlink control channel}
\newacronym{pdcp}{PDCP}{packet data convergence protocol}
\newacronym{pdf}{PDF}{probability density function}
\newacronym{pdp}{PDP}{power delay profile}
\newacronym{pdsch}{PDSCH}{physical downlink shared channel}
\newacronym{pe}{PE}{processing element}
\newacronym{peb}{PEB}{positioning error bound}
\newacronym{pef}{PEF}{Product Environmental Footprint}
\newacronym{per}{PER}{packet error rate}
\newacronym{pet}{PET}{privacy enhancing technology}
\newacronym{peet}{PET}{Polyethylene terephthalate}
\newacronym{pg}{PG}{path gain}
\newacronym{pgd}{PGD}{proximal gradient descent}
\newacronym{phy}{PHY}{physical layer}
\newacronym{pin}{PIN}{positive-intrinsic-negative}
\newacronym{pki}{PKI}{public key infrastructure}
\newacronym{pl}{PL}{path loss}
\newacronym{pla}{PLA}{physically large array}
\newacronym{pla2}{PLA}{polylactic acid}
\newacronym{pll}{PLL}{phase-locked loop}
\newacronym[plural=PMs,firstplural=person months (PMs)]{pm}{PM}{person month}
\newacronym{pmf}{PMF}{polymer microwave fiber}
\newacronym{pmic}{PMIC}{power management integrated circuit}
\newacronym{pmu}{PMU}{power management unit}
\newacronym{pn}{PN}{pseudo-noise}
\newacronym{po}{PO}{phase offset}
\newacronym{poc}{PoC}{proof of concept}
\newacronym{poe}{PoE}{power-over-Ethernet}
\newacronym{pps}{1PPS}{pulse per second}
\newacronym{pr}{PR}{phase reversal}
\newacronym{prb}{PRB}{Physical Resource Block}
\newacronym{prbs}{PRBs}{Physical Resource Blocks}
\newacronym{prs}{PRS}{Peripheral Reflex System}
\newacronym{ps}{PS}{Processing System}
\newacronym{psd}{PSD}{power spectral density}
\newacronym{pse}{PSE}{power sourcing equipment}
\newacronym{psk}{PSK}{phase shift keying}
\newacronym{psm}{PSM}{power saving mode}
\newacronym{pss}{PSS}{primary synchronisation signal}
\newacronym{ptp}{PTP}{precision-time protocol}
\newacronym{ptrs}{PTRS}{Phase-Tracking Reference Signals}
\newacronym{ptw}{PTW}{paging time window}
\newacronym{pv}{PV}{photovoltaic}
\newacronym{pw}{PW}{planar wavefront}
\newacronym{pwm}{PWM}{pulse width modulation}
\newacronym{qam}{QAM}{quadrature amplitude modulation}
\newacronym{qos}{QoS}{quality-of-service}
\newacronym{qpsk}{QPSK}{quadrature phase-shift keying}
\newacronym{qrrls}{QR-RLS}{QR decomposition based recursive least squares}
\newacronym{quadriga}{QuaDRiGa}{QUAsi Deterministic RadIo channel GenerAtor}
\newacronym{r2d}{R2D}{reader-to-device}
\newacronym{ra}{RA}{Random Access}
\newacronym{ram}{RAM}{random-access memory}
\newacronym{ran}{RAN}{radio access network}
\newacronym{rar}{RAR}{Random Access Response}
\newacronym{rat}{RAT}{radio access technology}
\newacronym{raw}{RAW}{restricted access window}
\newacronym{rb}{Rb}{Rubidium}
\newacronym{rbs}{RBS}{radio base station}
\newacronym{rbw}{RBW}{resolution bandwidth}
\newacronym{rc}{RC}{raised-cosine}
\newacronym{rcs}{RCS}{radar cross section}
\newacronym{rdl}{RDL}{redistribution layer}
\newacronym{re}{RE}{radio element}
\newacronym{red}{RED}{radio equipment directive}
\newacronym{redcap}{RedCap}{reduced capability}
\newacronym{relu}{ReLU}{rectified linear unit}
\newacronym{rf}{RF}{radio frequency}
\newacronym{rfeh}{RF-EH}{radio frequency energy harvesting}
\newacronym{rfic}{RFIC}{radio-frequency integrated circuit}
\newacronym{rfid}{RFID}{radio frequency identification}
\newacronym{rfpt}{RFPT}{radio frequency power transfer}
\newacronym{rfsoc}{RFSoC}{Radio Frequency System-on-Chip}
\newacronym{rfwpt}{RF-WPT}{radio frequency wireless power transfer}
\newacronym{rir}{RIR}{room impulse response}
\newacronym{ris}{RIS}{reflective intelligent surface}
\newacronym{rl}{RL}{reinforcement learning}
\newacronym{rlc}{RLC}{Radio Link Control}
\newacronym{rllmtc}{RLLMTC}{reliable low latency machine type communication}
\newacronym{rls}{RLS}{recursive least squares}
\newacronym{rms}{RMS}{root-mean-square}
\newacronym{rmse}{RMSE}{root-mean-square error}
\newacronym{rmt}{RMT}{random matrix theory}
\newacronym{rof}{RoF}{radio-over-fiber}
\newacronym{ros}{ROS}{robot operating system}
\newacronym{rpi}{RPi}{Raspberry Pi}
\newacronym{rpl}{RPL}{Routing Protocol for Low power and Lossy Networks}
\newacronym{rrc}{RRC}{Radio Resource Connection}
\newacronym{rreq}{RREQ}{route request packet}
\newacronym{rsrp}{RSRP}{Reference Signals Received Power}
\newacronym{rsrq}{RSRQ}{Reference Signal Received Quality}
\newacronym{rss}{RSS}{received signal strength}
\newacronym{rssi}{RSSI}{received signal strength indicator}
\newacronym{rtc}{RTC}{real time clock}
\newacronym{rtf}{RTF}{reader talks first}
\newacronym{rtk}{RTK}{real time kinematics}
\newacronym{rts}{RTS}{ray tracing simulator}
\newacronym{rtwt}{rTWT}{restricted TWT}
\newacronym{ru}{RU}{Radio Unit}
\newacronym{ruc}{rUC}{representative Use Cases}
\newacronym{rv}{RV}{random variable}
\newacronym{rw}{RW}{RadioWeaves}
\newacronym{rx}{RX}{receiver}
\newacronym{rzf}{RZF}{regularized zero forcing}
\newacronym{s-parameter}{S-parameter}{scattering parameter}
\newacronym{sa}{SA}{system aspects}
\newacronym{sa5}{SA}{Stand Alone}
\newacronym{sar}{SAR}{specific absorption rate}
\newacronym{sbl}{SBL}{sparse Bayesian learning}
\newacronym{sc}{SC}{single carrier}
\newacronym{scomp}{SC}{Split Computing}
\newacronym{scfde}{SC-FDE}{single-carrier modulation with frequency-domain-equalization}
\newacronym{scfdma}{SCFDMA}{single-carrier frequency division multiple access}
\newacronym{scs}{SCS}{sub-carrier spacing}
\newacronym{sdap}{SDAP}{service data adaptation protocol}
\newacronym{sdg}{SDG}{Sustainable Development Goal}
\newacronym{sdm}{SDM}{sigma-delta modulator}
\newacronym{sdma}{SDMA}{spatial-division multiple access}
\newacronym{sdn}{SDN}{software-defined network}
\newacronym{sdof}{SDoF}{sigma-delta over fiber}
\newacronym{sdr}{SDR}{software-defined radio}
\newacronym{se}{SE}{spectral efficiency}
\newacronym{sei}{SEI}{specific emitter identification}
\newacronym{ser}{SER}{symbol-error rate}
\newacronym{sf}{SF}{spreading factor}
\newacronym{sfn}{SFN}{single frequency network}
\newacronym{sfo}{SFO}{sampling frequency offset}
\newacronym{sfp}{SFP}{small form-factor pluggable}
\newacronym{sha}{SHA}{Secure Hash Algorithm}
\newacronym{SigMF}{SigMF}{Signal Metadata Format}
\newacronym{sdo}{SDO}{Standards Development Organization}
\newacronym{simo}{SIMO}{single-input multiple-output}
\newacronym{sinr}{SINR}{signal-to-interference-plus-noise ratio}
\newacronym{siso}{SISO}{single-input single-output}
\newacronym{slam}{SLAM}{simultaneous localization and mapping}
\newacronym{slc}{SLC}{spatial leakage suppression}
\newacronym{slerp}{SLERP}{spherical linear interpolation}
\newacronym{sma}{SMA}{SubMiniature version A}
\newacronym{smc}{SMC}{specular multipath component}
\newacronym{smps}{SMPS}{switched mode power supply}
\newacronym{smt}{SMT}{Surface Mount Technology}
\newacronym{sndr}{SNDR}{signal-to-noise-and-distortion ratio}
\newacronym{snidr}{SNIDR}{signal-to-noise-and-interference-and-distortion ratio}
\newacronym{snir}{SNIR}{signal-to-interference-plus-noise ratio}
\newacronym{snr}{SNR}{signal-to-noise ratio}
\newacronym{soc}{SoC}{state of charge}
\newacronym{SoC}{SOC}{System on Chip}
\newacronym{sota}{SotA}{state of the art}
\newacronym{sp}{SP}{service point}
\newacronym{spdt}{SPDT}{single pole double throw}
\newacronym{spi}{SPI}{Serial Peripheral Interface}
\newacronym{spst}{SPST}{single pole single throw}
\newacronym{sram}{SRAM}{static random-access memory}
\newacronym{srd}{SRD}{short-range device}
\newacronym{srls}{SRLS}{standard recursive least squares}
\newacronym{srs}{SRS}{Sounding Reference Signal}
\newacronym{ssb}{SSB}{synchronisation signal block}
\newacronym{ssd}{SSD}{solid state drive}
\newacronym{ssq}{SSQ}{simulator sickness questionnaire}
\newacronym{sta}{STA}{station}
\newacronym{steam}{STEAM}{science, technology, engineering, the arts, and mathematics}
\newacronym{svd}{SVD}{singular value decomposition}
\newacronym{sw}{SW}{spherical wavefront}
\newacronym{swipt}{SWIPT}{simultaneous wireless information and power transfer}
\newacronym{synce}{SyncE}{Synchronous Ethernet}
\newacronym{ta}{TA}{timing advance}
\newacronym{tau}{TAU}{tracking area update}
\newacronym{tcer}{TCER}{transported to consumed energy ratio}
\newacronym{tcp}{TCP}{Transmission Control Protocol}
\newacronym{tcxo}{TCXO}{temperature compensated crystal oscillator}
\newacronym{tdce}{TD-CE}{time-domain compression and expansion}
\newacronym{tdd}{TDD}{time division duplexing}
\newacronym{tdma}{TDMA}{time division-multiple access}
\newacronym{tdoa}{TDOA}{time-difference-of-arrival}
\newacronym{teg}{TEG}{Thermoelectric Generator}
\newacronym{tsg}{TSG}{Technical Specification Group}
\newacronym{thz}{THz}{Terahertz}
\newacronym{tinyml}{TinyML}{Tiny Machine Learning}
\newacronym{tinytl}{TinyTL}{Tiny Transfer Learning}
\newacronym{tinyol}{TinyOL}{Tiny Online Learning}
\newacronym{tinyrl}{TinyRL}{Tiny Reinforcement Learning}
\newacronym{tl}{TL}{Transfer Learning}
\newacronym{to}{TO}{timing offset}
\newacronym{toa}{TOA}{time-of-arrival}
\newacronym{tof}{ToF}{time-of-flight}
\newacronym{tosm}{TOSM}{through-open-short-match}
\newacronym{tpms}{TPMS}{Tire-Pressure Monitoring System}
\newacronym{trl}{TRL}{technology readyness level}
\newacronym{trp}{TRP}{Transmission Reception Point}
\newacronym{teng}{TENG}{Triboelectric nanogenerators}
\newacronym{tsn}{TSN}{time-sensitive networking}
\newacronym{ttf}{TTF}{tag talks first}
\newacronym{ttff}{TTFF}{Time To First Fix}
\newacronym{tti}{TTI}{transmission time interval}
\newacronym{ttm}{TTM}{time to market}
\newacronym{ttn}{TTN}{The Things Network}
\newacronym{tx}{TX}{transmitter}
\newacronym{twt}{TWT}{Target Wake Time}
\newacronym{uart}{UART}{Universal Asynchronous Receiver/Transmitter}
\newacronym{uav}{UAV}{unmanned aerial vehicle}
\newacronym{uc}{UC}{use case}
\newacronym{ucie}{UCIe}{Universal Chiplet Interconnect Express}
\newacronym{udp}{UDP}{User Datagram Protocol}
\newacronym{ue}{UE}{user equipment}
\newacronym{ugv}{UGV}{unmanned ground vehicle}
\newacronym{uhd}{UHD}{USRP hardware driver}
\newacronym{uhf}{UHF}{ultra-high frequency}
\newacronym{ul}{UL}{uplink}
\newacronym{ula}{ULA}{uniform linear array}
\newacronym{ule}{ULE}{ultra low energy}
\newacronym{ulp}{ULP}{ultra-low power}
\newacronym{uMIMO}{$\mu$-MIMO}{ultra-massive Multiple-Input Multiple-Output}
\newacronym{ummtc}{umMTC}{ultra massive machine type communication}
\newacronym{un}{UN}{United Nations}
\newacronym{unow}{uNOW}{unified non-orthogonal waveform}
\newacronym{upa}{UPA}{uniform planar array}
\newacronym{up}{UP}{ultra passive}
\newacronym{ura}{URA}{uniform rectangular array}
\newacronym{urllc}{URLLC}{ultra-reliable low-latency communications}
\newacronym{usrp}{USRP}{universal software radio peripheral}
\newacronym{uv}{UV}{unmanned vehicle}
\newacronym{uw}{UW}{unique word}
\newacronym{uwb}{UWB}{ultrawideband}
\newacronym{uwofdm}{UW-DFT-s-OFDM}{unique word DFT-s-OFDM}
\newacronym{v2v}{V2V}{vehicle-to-vehicle}
\newacronym{vco}{VCO}{voltage-controlled oscillator}
\newacronym{vep}{VEP}{virtual edge platform}
\newacronym{vlc}{VLC}{visible light communication}
\newacronym{vlp}{VLP}{visible light positioning}
\newacronym{vna}{VNA}{vector network analyzer}
\newacronym{voc}{VOC}{Voltatile Organic Compound}
\newacronym{votable}{VOTable}{Virtual Observatory Table}
\newacronym{vr}{VR}{virtual reality}
\newacronym{vuca}{VUCA}{volatile, uncertain, complex and ambiguous}
\newacronym{vswr}{VSWR}{Voltage Standing Wave Ratio}
\newacronym{wifi}{Wi-Fi}{Wireless Fidelity}
\newacronym{wirelesshart}{WirelessHART}{Wireless Highway Addressable Remote Transducer Protocol}
\newacronym{wur}{WuR}{wake-up radio}
\newacronym{wus}{WuS}{wake-up signal}
\newacronym{wb}{WB}{wideband}
\newacronym{wban}{WBAN}{wireless body area network}
\newacronym{wd}{WD}{Wireless distance}
\newacronym{wimax}{WiMAX}{Worldwide Interoperability for Microwave Access}
\newacronym{wlan}{WLAN}{wireless LAN}
\newacronym[plural=WPs,firstplural=work packages (WPs)]{wp}{WP}{work package}
\newacronym{wpc}{WPC}{Wireless Power Consortium}
\newacronym{wpt}{WPT}{wireless power transfer}
\newacronym{wr}{WR}{White Rabbit}
\newacronym{wrsn}{WRSN}{wireless rechargeable sensor network}
\newacronym{wsn}{WSN}{Wireless Sensor Network}
\newacronym{wcma}{WCMA}{weather conditioned moving average}
\newacronym{xets}{XETS}{cross exponentially tapered slot}
\newacronym{xlmimo}{XL-MIMO}{extremely large-scale MIMO}
\newacronym{xr}{XR}{extended reality}
\newacronym{z3ro}{Z3RO}{zero third-order distortion}
\newacronym{zed}{ZED}{Zero-Energy device}
\newacronym{zf}{ZF}{zero-forcing}
\newacronym{zmcscg}{ZMCSCG}{zero mean circularly symmetric complex Gaussian}
\newacronym{zmq}{ZMQ}{ZeroMQ}
\newacronym{ztofdm}{ZT-DFT-s-OFDM}{zero-tail DFT-s-OFDM}

\newacronym{mobc}{MoBC}{monostatic backscatter communication}

\newacronym{bibc}{BiBC}{bistatic backscatter communication}
\newacronym{t2t}{T2T}{tag-to-tag}
\newacronym{pls}{PLS}{physical layer security}

\newacronym{cea}{CEA}{controlled environment agriculture}
\newacronym{abd}{ABD}{ambient-backscattering device}
\newacronym{keh}{KEH}{kinetic energy harvester}
\newacronym{svm}{SVM}{support vector machine}
\newacronym{wisp}{WISP}{Wireless Identification and Sensing Platform} 
\glsdisablehyper

\begin{document}

\title{Energy-Neutral Coverage Optimization by Joint Deployment and Scheduling in Ambient IoT Devices with Directional Sensing
}
\author{David E. Ru\'{i}z-Guirola,~\IEEEmembership{Member,~IEEE,}
        Samuel~Montejo-S\'{a}nchez,~\IEEEmembership{Senior Member, IEEE}, 
        Richard~Demo~Souza,~\IEEEmembership{Senior Member, IEEE},  and~Onel~L.~A.~L\'{o}pez,~\IEEEmembership{Senior Member, IEEE}
\thanks{David E. Ru\'{i}z-Guirola and Onel L. A. L\'{o}pez are with the Centre for Wireless Communications, University of Oulu, Finland. \{David.RuizGuirola, Onel.AlcarazLopez\}@oulu.fi. 
Samuel Montejo-S\'{a}nchez is with the {Instituto Universitario de Investigaci\'{o}n y Desarrollo Tecnol\'{o}gico, Universidad Tecnol\'{o}gica Metropolitana}, Santiago, Chile. \{smontejo@utem.cl\}.
Richard Demo Souza is with the Federal University of Santa Catarina, Florianópolis, SC, Brazil. \{richard.demo@ufsc.br\}} 
\thanks{This work has been partially supported by the Research Council of Finland (Grants 369116 (6G Flagship Programme) and 362782 (ECO-LITE)), the Finnish Foundation for Technology Promotion, and the European Commission through the Horizon Europe/JU SNS project AMBIENT-6G (Grant 101192113), in Chile by ANID FONDECYT Regular 1241977 and ANID CPS-RTC CIA250016, and in Brazil by CNPq (317368/2025-7) and RNP/MCTI Brasil 6G (01245.020548/2021-07).}
}

\maketitle

\begin{abstract}
\Gls{aiot} devices rely on energy harvesting and duty cycling to sustain operation, thereby fundamentally changing collaborative sensing compared with traditional always-ON sensor networks. In this paper, we study the joint deployment and sensing scheduling of \gls{aiot} devices equipped with directional sensing. We explore four solution strategies: (i) a grid deployment with static duty cycling, (ii) a centralized policy-gradient \gls{rl} approach that begins with a grid deployment and learns energy-aware device relocation and duty-cycling policies, (iii) a mixed-integer \gls{lp} approach that couples static deployment design with duty-cycle allocation, and (iv) a hybrid \gls{lp}+\gls{rl} that combines optimization-based initialization with learning-based refinement. Using representative \gls{aiot} use cases, we evaluate coverage as a function of device density, field-of-view, and maximum feasible duty cycle, determined by harvested energy and device consumption. Numerical results indicate that the proposed \gls{lp}+\gls{rl} and \gls{rl} policies consistently outperform both the grid baseline and the \gls{lp}-based method, achieving up to 2x higher mean effective coverage in low and medium {\gls{eh}} regimes. In contrast, the standalone \gls{lp} method remains limited by its conservative static duty cycle allocation under tight \gls{eh} constraints. Moreover, the structured initialization of the \gls{lp}+\gls{rl} method substantially accelerates convergence, reducing the total offline optimization {time by up to 10x compared to the standalone \gls{rl}.} 


\end{abstract}


\glsresetall 

\IEEEpeerreviewmaketitle


\section{Introduction}\label{sec:intro}
\IEEEPARstart{S}{ustainable}
development {requires} technological solutions that effectively balance economic growth, social equity, and environmental integrity. {This includes} optimizing resource management and enhancing access to essential services, including smart healthcare, smart transportation, and smart cities~\cite{lopez2023energy}. The \gls{iot} plays a crucial role in this process by connecting billions of sensor devices that collect data on various phenomena such as temperature, pressure, motion, mobility, and safety. This data is then transformed into actionable insights for autonomous control~\cite{ruiz2026context}.  
{However, with} projections exceeding 38 billion \gls{iot} devices connected by 2029, the prevailing reliance on primary (non-rechargeable) batteries is increasingly incompatible with sustainability and scalability, driving frequent maintenance and higher lifecycle costs~\cite{ericsson2023}. This is why energy-efficient operations and extending the devices’ lifetime have been at the center of attention in recent years. 

As deployments grow and underpin critical infrastructure, stringent quality-of-service demands such as reliability, latency, and energy lifetime intensify. Devices are evolving from passive data reporters to coordinated agents, which increases the value of the data and adds complexity to life-cycle management and control~\cite{ruiz2026context}. In this context, \gls{aiot} envisions large-scale deployments of \gls{iot} devices that 
rely on 
\gls{eh}. 
\gls{aiot} devices may operate at $\mu$W-level average power and lower, waking up only sporadically to sense, process, and report {events}~\cite{3gppTS38391,famaey2026survey}. This approach results in decades-long energy autonomy, thereby reducing maintenance costs and extending device {lifetime}~\cite{famaey2026survey, lopez2024zero}. Note that continuous monitoring, however, cannot be supported by a single \gls{aiot}, calling for collaborative sensing approaches where multiple devices jointly provide sufficient spatial and temporal coverage of a region of interest~\cite{he2022collaborative}.

Previous studies have focused on enhancing sensor deployment and coverage in the \gls{iot} sensing layer. For example, authors in~\cite{wang2024method} optimize node positions and orientations to avoid coverage holes and maximize the effective coverage area. In industrial \gls{iot} systems, joint optimization of coverage and connectivity has been shown to improve data quality while minimizing the number of sensors and maintaining connectivity in~\cite{dou2015optimizing}. Additionally, policy-driven deployments along with sleep/wake scheduling have proven effective in extending network lifetime. These solutions have been achieved through centralized/distributed~\cite{zhu2018optimal} heuristics, and more recently, through \gls{rl} techniques, particularly for solar-powered \gls{iot} networks~\cite{he2022collaborative}. 


However, existing approaches~\cite{he2022collaborative,dou2015optimizing,wang2024method} assume continuously active sensors or relatively dense duty cycles, and mostly focus on battery-powered nodes with isotropic sensing models. As a result, they fail to integrate \gls{eh} constraints, \gls{fov} heterogeneity, and sensing schedules in \gls{aiot} settings. Although maximizing coverage with the minimum number of devices has been explored previously, \textit{e.g.,}~\cite{nematzadeh2023maximizing,tossa2022area,11251244}, achieving the same coverage with \gls{eh}‑constrained \gls{aiot} devices that do not have continuous or predictable energy availability is substantially more challenging.  
This is especially relevant in \gls{aiot} deployments, where devices may have limited or directional \gls{fov} and therefore must coordinate to cover large areas under strict energy budgets.

Depending on the sensing modality and application, \gls{aiot} devices may exhibit narrow (\textit{e.g.}, $30^\circ–60^\circ$), panoramic ($120^\circ–180^\circ$), or omnidirectional ($360^\circ$) \gls{fov}. Representative \gls{fov} ranges and applications are summarized in Table~\ref{tab:FoV_classes}. 
The \gls{3gpp} envisions potential service requirements for \gls{aiot}-enabled applications, including sensing in smart homes, smart agriculture, and collaborative sensing, with devices expected to operate maintenance-free for more than 10 years. Therefore, to maintain energy‑neutral operation, a device’s duty cycling policy must ensure that expected energy consumption over an interval $T$ does not exceed the expected energy harvested during $T$. The horizon $T$ should be chosen in relation to the device’s storage capacity so that transient energy deficits do not render the device unavailable. 
Under these conditions, sustaining performance while minimizing energy consumption requires a reliable policy for determining when to sleep and when to sense~\cite{ruiz2026energy}. 
\begin{table}[t]
  \caption{Representative \gls{fov} classes, sensing modalities, and applications}
  \label{tab:FoV_classes}
  \centering
  \begin{tabular}{p{2cm}p{1cm}p{4.6cm}}
    \hline
    \textbf{Sensor type} & \textbf{\gls{fov}} & \textbf{Representative use cases~\cite{3gpp_tr22840}} \\
    \hline
     Infrared presence sensor / PIR 
        & $60^\circ$~\cite{sirmacek2020occupancy}
        & Smart-home presence sensing, gas/fire alert support, occupancy-aware heating, ventilation, and lighting control, and localized indoor monitoring. \\
    Panoramic intruder detector; fisheye camera
        & $180^\circ$~\cite{wakai2021deep} 
        & Smart-campus safety, indoor surveillance, museum guide and exhibition monitoring, and medical or asset facility supervision/assistance. \\
    Omnidirectional sensors; 360$^\circ$ LiDAR/radar (isotropic \gls{fov})
        & $360^\circ$~\cite{ruiz2026context}
        & Automated warehousing, hospital instrument inventory and positioning, airport/shipping asset tracking, personal-belongings finding, and smart-agriculture or warehouse-scale monitoring. \\
    \hline
  \end{tabular}
\end{table}

Herein, we study coverage optimization in \gls{aiot} networks where the devices collaborate to monitor a region. 
While~\cite{ruiz2026energy} optimizes duty cycling and sensing scheduling for a fixed deployment to reduce energy consumption, {our} work jointly optimizes device deployment and sensing operation under an explicit energy-consumption constraint to ensure long-term device availability and sustainable network reliability. 
We focus on energy-efficient sensing through coordinated sleep/wake scheduling and deployment decisions that maximize spatio‑temporal coverage under per‑device energy budgets. 
The main contributions of this paper are as follows:
\begin{itemize}
  \item We formulate a long-term coverage maximization problem under energy-neutral operation. 
  Unlike prior work that separately considers static coverage and deployment~\cite{he2022collaborative,dou2015optimizing,wang2024method,nematzadeh2023maximizing,tossa2022area,11251244,zhu2018optimal} or energy management for fixed deployments~\cite{ruiz2026energy}, our formulation jointly optimizes device placement and temporal sensing schedules under intermittent energy availability for representative \gls{aiot} use cases.    
  \item We propose and compare three energy-aware optimization frameworks: (i) an \gls{lp}-based method for joint deployment and duty-cycle allocation; (ii) a grid-initialized \gls{rl} policy that learns device relocation and sensing decisions; and (iii) a hybrid \gls{lp}+\gls{rl} approach that uses the \gls{lp} solution to initialize and accelerate the learning stage. We analyze their computational complexity and runtime, showing that the hybrid method converges faster than standalone \gls{rl}. 
  \item We evaluate the proposed methods' coverage as a function of device density, \gls{fov}, and maximum feasible duty cycle. We show that the \gls{rl} policy and, in particular, the hybrid \gls{lp}+\gls{rl} method provide consistently the best coverage performance across device densities, \glspl{fov}, and feasible duty-cycle regimes, especially under low and medium energy budgets.
\end{itemize}




The rest of the paper is organized as follows. Section~\ref{sec:2_system} describes the system model, including the event sensing modeling, the energy consumption and harvesting model, and the coverage metric assessment. Section~\ref{sec:problem_formulation} formulates the optimization problem for the duty cycling and wake-up setup, while Section~\ref{sec:policy} presents solutions based on heuristics, Voronoi diagrams, \gls{lp}, and \gls{rl}. The proposed methods are evaluated through numerical simulations in Section~\ref{sec:results}. Finally, we conclude the paper in Section~\ref{sec:conclusions}. 


\section{System model}\label{sec:2_system}

\begin{figure}[t!]
\centerline{\includegraphics[width=\columnwidth]{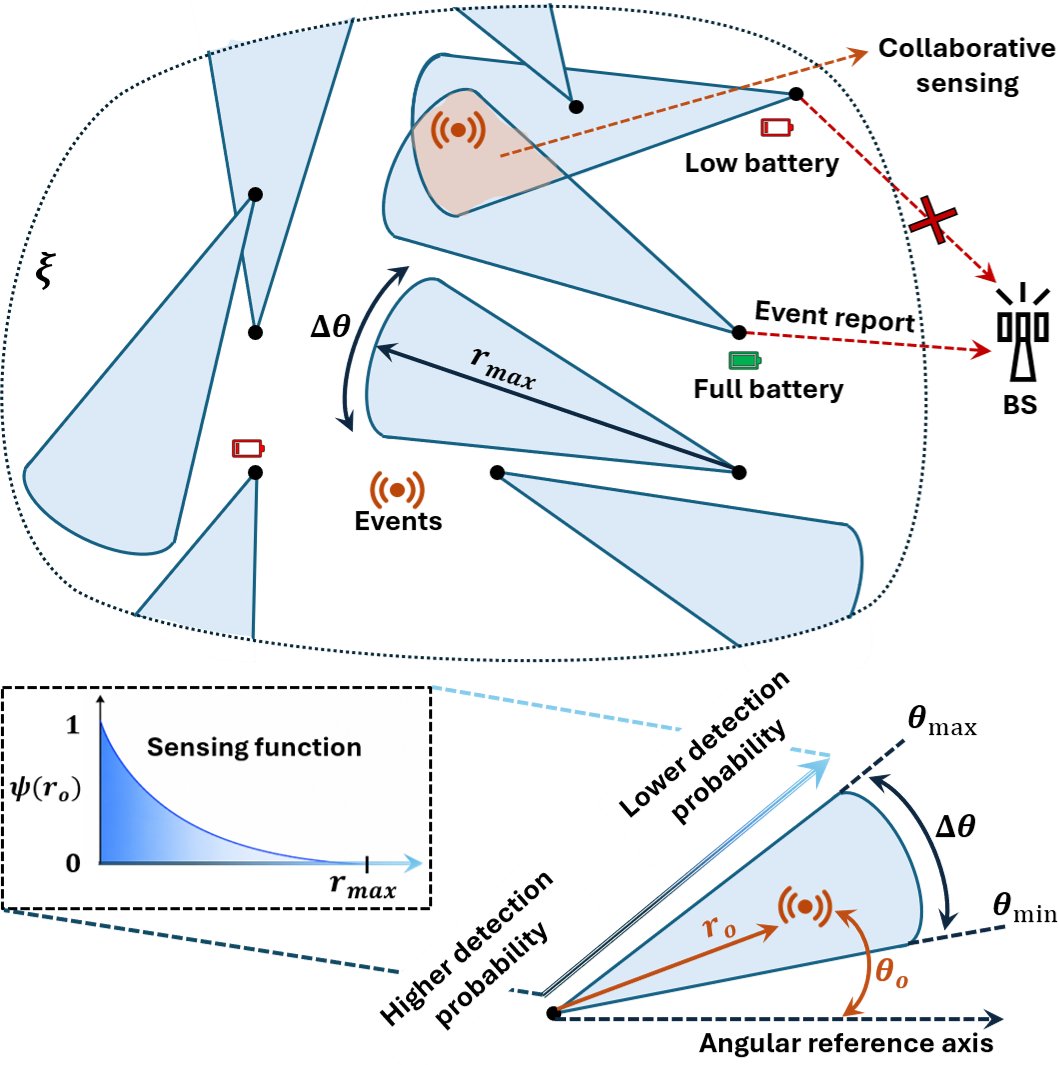}}
\vspace{-2mm}
\caption{\gls{aiot} system model with collaborative sensing and \gls{fov} constraints. A BS coordinates a set of \gls{eh} \gls{aiot} devices deployed over a region of interest $\xi$. Each device has a finite energy buffer and a directional sensing modality with \gls{fov} width $\Delta\theta = \theta_{\max}-\theta_{\min}$ and maximum sensing radius $r_{\max}$. An event can be detected only if a device is (i) active (i.e., awake and not in sleep mode), (ii) located within the sensing radius $r_{max}$, and (iii) oriented so that the event lies inside its \gls{fov} $\Theta$. The shaded blue sector in the sensing function illustrates the distance-dependent sensing decay, with higher detection reliability near the device and lower reliability toward $r_{\max}$. 
}
\vspace{-3mm}
\label{fig:system_model}
\end{figure}


Consider an \gls{aiot} scenario where a \gls{bs}, or an assisting node, acts as the gateway for a set ${\mathcal{N}}$ of $N$ \gls{aiot} devices, as illustrated in \figurename~\ref{fig:system_model}. An A-IoT $j\in\mathcal{N}$ is located at coordinates $(x_j,y_j)$. The devices are deployed to monitor processes within the coverage region $\xi \in \mathbb{R}^2$. When an event occurs, it triggers the devices to send information about the event to the \gls{bs}. 
Every device is assumed to have identical \gls{fov}, an \gls{eh} module, and an energy buffer, along with application-specific hardware that includes a \gls{rf} communication module for necessary communication with the BS and a low-power microcontroller. Furthermore, time is divided into discrete \glspl{tti} with duration $\tau$, indexed by $t=1,\dots, T$. 

We consider an event-driven scenario, wherein the device operates under duty cycling, getting active based on proximity to critical events, such as a moving object in motion detection applications or in a fire-alarm system, and transmitting data to the BS upon detection.  
At each \gls{tti}, an event is generated with probability $p_{\mathrm{ev}}$, according to a Bernoulli arrival process, while its  location is assumed to be uniformly distributed over the monitored region~\cite{ruiz2026energy}. Note that, although this event model is adopted for performance evaluation, the proposed framework is not tied to a specific event distribution and can be extended to other spatial or temporal event models. Moreover, it is assumed that a device without energy in storage cannot perform sensing and reporting. Additionally, \gls{aiot} devices can harvest energy whenever it is available, regardless of their operational state.

\subsection{Event sensing power function modeling}\label{Sec:2.1}


The \gls{fov} of a device is defined as the angular sector it can sense. Each device is characterized by a sensing \gls{fov} $\Theta=[\theta_{\min},\theta_{\max}]$, with angular width  $\Delta\theta=\theta_{\max} - \theta_{\min}$. Accordingly, the angular detection function is given by 
\begin{equation}\label{eq:angular_detection}
I(\theta_o) =
\begin{cases}
1, & \text{if } \theta_o \in \Theta,\\
0, & \text{otherwise}.
\end{cases}
\end{equation} 
Thus, objects located at an angle $\theta_o$ within the range $\Theta$ can be detected, whereas those outside this range cannot.  
Conditioned on the event falling inside the device’s \gls{fov}, we model the sensing capability as a distance-dependent function that decays with the distance up to a maximum range $r_{\max}$. 
This model captures the practical behaviour of directional sensors, whose detection reliability typically decreases with range due to signal attenuation and reduced measurement strength~\cite{sirmacek2020occupancy,wakai2021deep}.
For a given \gls{tti}, the sensing function is given by
\begin{equation}
\psi(r_o) =
\begin{cases}
\exp(-\eta r_o), & \text{if } r_o \in [0, r_{\max}],\\
0, & \text{otherwise},
\end{cases}
\end{equation}
where $\eta$ controls the decay rate. Combining the angular and distance components, the sensing function over one \gls{tti} is given by
\begin{equation}\label{eq:sensing_function}
p(r_o, \theta_o) = \psi(r_o) I(\theta_o).
\end{equation}
If no device detects the event during the \gls{tti} in which it occurs, the detection task is considered unsuccessful.

\subsection{Energy consumption and harvesting model}\label{sec:EC_EH}

We model the energy dynamics of each \gls{aiot} device in discrete time, indexed by the \gls{tti} $t$. Each device is equipped with an energy harvester and a finite-capacity battery with maximum stored energy $E_{\max}${, such that} $B_j(t)\in[0,E_{\max}]$ denotes the battery energy of device $j$ at time $t$.

The energy consumption associated with sensing and reporting is modeled through the per-slot consumption variable $E_j^{\mathrm{C}}(t)$. In particular, when device $j$ is active and performs a sensing (reporting) operation, it consumes a fixed energy amount $E_{\mathrm{sense}}$ ($E_{\mathrm{Tx}}$). Otherwise, its sensing-related energy consumption is assumed to be zero. Hence,
    $E_j^{\mathrm{C}}(t)\in\{0,E_{\mathrm{sense}},E_{\mathrm{Tx}}\}.$
If the available battery energy is insufficient to support sensing or reporting, \textit{i.e.}, ${B_j(t)<E_{\mathrm{sense}}+E_{\mathrm{Tx}}}$, then the corresponding operation cannot be successfully executed.

Let $E_j^{\mathrm{H}}(t)$ denote the energy harvested by device $j$ during slot $t$. We assume an i.i.d. binary {\gls{eh}} process at \gls{tti} $t$, $\{E_j^{\mathrm{H}}(t)\}_{t=1}^{\infty}$, with $\lambda^H_j$ denoting the average energy-arrival rate of the harvester for device $j$. When active, the harvester delivers a fixed quantum $E_\mathrm{H}$ units to the device; otherwise, it provides $0$ units,
    $E_j^{\mathrm{H}}(t)\in\{0,E_{\mathrm{H}}\}$. 
This stochastic \gls{eh} model provides a simple and non-causal representation of the energy inflow process while remaining general enough to capture ON/OFF harvesters that deliver fixed energy quanta, such as time-slotted solar harvesting, periodic vibration harvesters in industrial environments, or dedicated light/RF energy sources~\cite{ku2015advances,gindullina2021age}. 
Accordingly, the battery evolution of device $j$ is given by
\begin{equation}
    B_j(t\!+\!1)\!=\! \min\!\Big\{\!E_{\max},\max\!\big[B_j(t)\!-\!E^{\mathrm{C}}_j(t),\,0\big]\!+\!E^{\mathrm{H}}_j(t)\Big\}.
\label{eq:battery_evolution}
\end{equation}

Finally, let $f_j\in[0,1]$ denote the duty cycle of device $j$. Under energy-neutral operation, the long-term average consumed energy cannot exceed the long-term average harvested energy. Thus, the maximum feasible duty cycle of device $j$ is
\begin{equation}\label{f_max}
    f_{j,\max}=\frac{\lambda^H_j E_{\mathrm{H}}}{E_{\mathrm{sense}} + E_{\mathrm{Tx}}}.
\end{equation}
This upper bound explicitly connects a device’s sensing activity to its \gls{eh} capacity and to the condition for battery‑sustainable operation.

\section{Problem formulation}\label{sec:problem_formulation}

We aim to {determine an optimal deployment that maximizes the coverage area of \gls{aiot} devices, enabling reliable detection of events in the region of interest at any given \gls{tti}. 
Once optimized, the device locations, \gls{fov} orientations, and duty-cycling remain fixed during network operation. However, the instantaneous sensing state of each device may vary across \glspl{tti} according to its predefined duty cycle and available harvested energy.}
Successful event detection depends on the device deployment, \gls{fov} orientations, sensing schedules, and available battery energy. Hence, the key challenge is to balance energy consumption to keep a sufficient number of devices properly positioned and oriented so as to reduce blind spots and improve event detectability. Since devices with insufficient energy cannot sense or report, these decisions must account for the temporal evolution of the stored energy.  

Let ${\boldsymbol{\mathcal{T}}(t) = [\mathcal{T}_1(t), \mathcal{T}_2(t), \dots, \mathcal{T}_N(t)]}$ denote the state of all devices at \gls{tti} $t$. The local state of device $j$ is defined as
\begin{equation}
  {\mathcal{T}}_j(t) = \big\{ \delta_j(t), \Theta_j, p_j(r_o,\theta_o),
                 \mathcal{D}_j, B_j(t) \big\},
  \label{eq:local_state}
\end{equation}
where $\mathcal{D}_j=\{(x_j,y_j)\}$ denotes the position of device $j$, and $\delta_j(t)\in\{0,1\}$ is the duty-cycling action (sleep/sense) at \gls{tti} $t$. The battery level $B_j(t)$ is an internal state that evolves according to the \gls{eh} and consumption dynamics in Section~\ref{sec:EC_EH}.

Thus, for a given deployment and sensing action, the sensing contribution of device $j$ to point $(x,y)$ is then
\begin{equation}
  q_j(x,y,t)=\delta_j(t)\,p_j(r_{(x,y)},\theta_{(x,y)}),
  \label{eq:qj}
\end{equation}
which is zero whenever the device is sleeping or the point lies outside its effective sensing region, as described in \eqref{eq:angular_detection}-\eqref{eq:sensing_function}. 
The probability that an event $o$ at a point $(x_o,y_o)$ is sensed by at least one active device at a \gls{tti} $t$ is given by
\begin{equation}
  s(x_o,y_o,t)=1-\prod_{j=1}^{N}\Big(1-q_j(x_o,y_o,t)\Big).
  \label{eq:sxy}
\end{equation} 

We focus on target and area coverage, where the latter is defined over the region of interest $\xi$. The monitored region is considered to be sensing-covered if every point $(x,y)\in\xi$ lies within the effective sensing region of at least one \gls{aiot} device. Analogously, the region is $m$-sensing-covered if every point lies within the effective sensing regions of at least $m$ \gls{aiot} devices. Then, the sensing coverage metric per \gls{tti}, denoted by `$\mathrm{Cov}(t)$', is defined as the spatial average sensing probability over the region $\xi$, {namely}
\begin{equation}
    \mathrm{Cov}(t) = \frac{1}{|\xi|}\int_{\xi} s(x_o,y_o,t)\,\mathrm{d}x\,\mathrm{d}y,
    \label{eq:cover}
\end{equation}
where $|\xi|$ is the Lebesgue measure (area) of the region $\xi$~\cite{burk1997lebesgue}. This metric provides a normalized measure of the network’s average ability to observe points in the monitored area, accounting for sensing probability rather than binary coverage. 
The instantaneous spatial coverage in~\eqref{eq:cover} depends on the network state $\boldsymbol{\mathcal{T}}(t)$ through the active-device set. Then, the long-term coverage optimization problem is given by
\begin{align}\label{eq:cov_opt}
  \max_{\mathcal{D},\,\Theta,\,\pi}\quad
  \frac{1}{T}\sum_{t=1}^{T} \mathbb{E}_{\pi}\!\left[\mathrm{Cov}(t)\right], 
\end{align}
where $\pi$ is a sensing policy that maps the observed state into duty-cycling actions $\boldsymbol{\delta}(t)$ and \gls{fov} $\Theta$. The expectation is subject to the battery dynamics in Section~\ref{sec:EC_EH} and energy availability constraints. Problem~\eqref{eq:cov_opt} therefore jointly determines an offline device deployment and a temporal sensing policy that maximizes time-average coverage while maintaining sustainable operation.


\section{Deployment policy}\label{sec:policy}

Even in a simplified static setting, problem~\eqref{eq:cov_opt} is computationally intractable. Consider a single \gls{tti} with fixed battery levels, no energy dynamics, and a constraint that at most $K<N$ devices can be active. Selecting the subset of devices (and their \gls{fov}) that maximizes $\mathrm{Cov}$ reduces to a maximum coverage problem, where each device corresponds to a sensing coverage (set of points it can cover) and the objective is to maximize the number of covered points. Maximum coverage is known to be NP-hard, as it generalizes the classical set-cover problem. When device positions $(x_j,y_j)$ are also optimization variables, the problem additionally includes a location/deployment component that is itself NP-hard~\cite{dou2015optimizing,he2022collaborative}.
{In this sense, population-based metaheuristics, such as genetic algorithms and particle swarm optimization, can provide sub-optimal solutions for the static deployment problem}~\cite{nematzadeh2023maximizing}. 
{However,} when the time dimension and \gls{eh}-driven battery dynamics are incorporated, \eqref{eq:cov_opt} becomes a control problem that is substantially more challenging than the static coverage formulations in~\cite{dou2015optimizing}. Therefore, realistic network sizes must resort to heuristics, relaxations, or learning-based approximations {that exploit the sequential structure introduced by time-varying battery states and stochastic \gls{eh} arrivals.}

To provide a strong and interpretable starting point, we initialize device locations on a regular grid covering the region of interest, as commonly used in policy-driven area coverage studies, \textit{e.g.,}~\cite{dou2015optimizing,he2022collaborative}. This grid serves both as a baseline for heuristic, grid‑based policies with static duty cycling and as an initial state for the learning‑based policy, which can iteratively refine device positions and sensing schedules.

Given an observed state $T(t)$, the \gls{bs} may adjust $\boldsymbol{\delta}(t) = [\delta_1(t),\dots,\delta_N(t)]$, $\boldsymbol{\theta}(t) = [\theta_1(t),\dots,\theta_N(t)]$, and the device positions $\{(x_j(t),y_j(t))\}_{j\in\mathcal{N}}$ according to a policy. 
Note that these position updates are virtual actions evaluated within the training environment and do not imply the devices to change their location in reality. This is an offline problem in which the final learned locations define a fixed deployment, while the sensing and duty-cycling decisions are subsequently applied during network operation. Thus, the \gls{bs} does not require real-time physical relocation feedback. 
In this work, we use a grid deployment as a baseline and propose three solution methods, which are explained in detail below. 

\subsection{Grid + static duty cycling}

Herein, the \gls{aiot} devices remain fixed at their initial grid locations and follow a pre-designed periodic active-sleep pattern $(t_{\text{on}},t_{\text{off}})$ that satisfies the device’s average \gls{eh} budget but does not adapt to instantaneous battery state or to transient coverage gaps. It is proven that, for always-ON devices, the asymptotically optimal deployment for covering a continuous convex region $\xi$ is a regular triangular pattern with inter-device spacing $\sqrt{3}\,r_{\max}$, regardless of the specific shape of $\xi$. Thus, the Voronoi polygon for each device is a regular hexagon with length $r_{\max}$~\cite{dou2015optimizing}. 

\subsection{\gls{rl}-based Proposal}\label{RL_method}

Devices are initially arranged on a grid, with the \gls{bs} acting as a centralized learning agent. The \gls{bs} iteratively refines both the positions of the devices and their duty-cycling/\gls{fov} decisions to enhance coverage while meeting \gls{eh} constraints. In this context, the centralized \gls{rl} algorithm optimizes deployment and sensing policies starting from the grid initialization. The goal is to obtain final device locations and sensing configurations that collectively maximize the long-term coverage metric. 

The proposed \gls{rl} approach is a centralized episodic policy-gradient method~\cite{agarwal2023reinforcement} in which the \gls{bs} learns a stochastic policy through categorical softmax distributions over position, \gls{fov} orientation, and duty-cycling. 
We model the device deployment and \gls{eh} processes as the environment, while the central controller at the \gls{bs} functions as the agent in a centralized learning framework. At each learning step, the agent observes the current state $\boldsymbol{\mathcal{T}(t)}$, takes an action (moving devices and/or updating $\boldsymbol{\delta}(t)$ and $\boldsymbol{\theta}(t)$), and receives a reward $r(t)$ that reflects both coverage and energy efficiency. We define the reward as
\begin{equation}
  r(t) = \Biggl( 1 - \frac{1}{N} \sum_{j=1}^N \delta_j(t) \Biggr) + \mu \, \mathrm{Cov}(t),
  \label{eq:reward}
\end{equation}
where $\mu > 0$ is a weighting factor\footnote{The value of $\mu$ must be selected carefully. An excessively large $\mu$ makes the energy-saving term negligible, compromising the long-term availability.}. The first term penalizes activating too many devices (reducing redundant overlap and saving energy), while the second term promotes high coverage availability. The goal is to determine an approximately optimal policy $\pi^\star$ that maximizes the long-term expected reward~\cite{yang2024learning,sheng2025beyond}
\begin{align}
  \pi^\star = \arg\max_\pi \;
  \mathbb{E}\left[ \sum_{t=1}^{T} \gamma^{t-1} r(t) \right],
\end{align}
with discount factor $\gamma \in (0,1]$. 
The policy parameters are then updated through a gradient rule to reduce variance during training. In practice, the \gls{rl} procedure converges to an optimized deployment and sensing schedule, which we then evaluate and compare against the grid-based heuristic baseline. 


\subsection{Convex Approach and Greedy Deployment}\label{sec:LP}

To obtain a tractable solution to \eqref{eq:cov_opt}, we discretize the region of interest into $P$ grid points and restrict device positions to a finite set of candidate locations $\mathcal{D}$ and \gls{fov} orientations $\Theta$. Each candidate configuration $m \in \{1,\dots, M\}$ corresponds to a location/orientation pair. Then, we compute a binary coverage matrix $a_{i,m}$ where $a_{i,m} = 1$ if candidate $m$ can sense grid point $i$ according to the sensing model in Section~\ref{Sec:2.1}, or $a_{i,m}=0$ otherwise.

A static deployment problem can then be written as a maximum-coverage mixed integer \gls{lp}~\cite{zhu2018optimal} \begin{subequations}
\begin{alignat}{4}
    \max_{x,u}& \quad
     \frac{1}{P}\sum_{i=1}^{P} u_i
    \\
    \text{s.t.}& \quad
    \hspace{12mm} u_i&& \leq \sum_{m=1}^{M} a_{i,m}x_m,
    \quad \forall i,
    \\
    &
    \sum_{m=1}^{M} c_m x_m&& \leq C_{\mathrm{budget}},
    \\
    &
    \hspace{12mm} x_m&& \in \{0,1\},
    \quad \forall m,
    \\
    &
    \hspace{12mm} u_i&& \in \{0,1\},
    \quad \forall i,
\end{alignat}
\label{LP_problem}
\end{subequations} 
\hspace{-1.8mm}where $x_m$ indicates whether a device is deployed with configuration candidate $m$, $u_i \in \{0,\,1\}$ indicates whether grid point $i$ is covered, $c_m$ is the cost associated with the configuration candidate $m$, and $C_{\text{budget}}$ is the energy budget. This formulation is NP-hard~\cite{he2022collaborative}, but {small instances with limited number of candidate configurations and discretized coverage points} can be solved using standard \gls{lp} solvers, whereas larger ones can be approximated using relaxations and greedy heuristics~\cite{zhu2018optimal}.

By relaxing the binary constraints to $x_m\in[0,1]$ and $u_i\in[0,1]$, the problem reduces to a \gls{lp}. {The relaxed solution $\{x_m^\star\}$ provides a priority score for each candidate configuration, as shown in lines~1-3 of} Algorithm~\ref{alg:greedy_deployment}. 
{The algorithm repeatedly evaluates every unselected candidate $m$ that satisfies the remaining budget constraint. Specifically, lines~7-11 compute the marginal gain $\Delta_u$, defined as the number of previously uncovered points that candidate $m$ would cover, and normalize it by the corresponding cost $c_m$. In each iteration, the feasible candidate with the largest marginal coverage gain per unit cost is selected, using the relaxed variable $\{x_m^\star\}$ to break ties (line~12). Finally, lines~13--15 update the selected set ($\mathcal{D}_j,\,\Theta_j$), covered-point set, and consumed budget. This process continues until no additional candidate can be selected without exceeding $C_{\mathrm{budget}}$. This budget restricts the feasible solution space by imposing an upper limit on the number of configurations that can be selected.} 
\begin{algorithm}[t]
\caption{Greedy deployment with \gls{lp} relaxation}
\label{alg:greedy_deployment}
\small
\begin{algorithmic}[1]

\State \textbf{Input:} $\mathcal{D}$, ${\Theta}$, $\{a_{i,m}\}$, $\{c_m\}$, and $C_{\mathrm{budget}}$

\State Candidate set $\mathcal{M}=\mathcal{D}\times\Theta$

\State Solve~\eqref{LP_problem}, with $x_m,u_i\in[0,1]$, to obtain $\{x_m^\star\}$

\State Initialize the selected set $\mathcal{S}\gets\emptyset$
\State Initialize the covered-point set $\mathcal{U}\gets\emptyset$
\State Initialize the used budget $C\gets0$

\While{at least one unselected candidate fits  $m\in\mathcal{M}\setminus\mathcal{S}$
such that $C+c_m\le C_{\mathrm{budget}}$}

    \For{each $m$}
        \State Compute $\Delta_u =\sum u_i :  \left\{a_{i,m}=1,\, i\setminus\mathcal{U}\right\}$
        \State Compute the coverage gain per unit cost,  ${\Delta_u}/{c_m}$
    \EndFor

    \State Select $m^\star\in\arg_m\max \, {\Delta_u}/{c_m}$, prioritizing devices with 
    \Statex \hspace{4.8mm}largest relaxed score $x_m^\star$
    

    \State $\mathcal{S}\gets\mathcal{S}\cup\{m^\star\}$
    \State $\mathcal{U}\gets
    \mathcal{U}\cup\{i:a_{i,m^\star}=1\}$
    \State $C\gets C+c_{m^\star}$

\EndWhile

\State \Return $\mathcal{S}$

\end{algorithmic}
\end{algorithm}

Given the resulting deployment, we then design long-term duty cycles for each device under the \gls{eh} constraints, $f_{j,\max}$. 
We approximate the coverage contribution of device $j$ by $\alpha_j = \sum_{i} a_{i,j}$, i.e., the number of grid points it can cover, and solve the \gls{lp}
\begin{subequations}
\begin{alignat}{4}
  \max_{\{f_j\}} \quad &
    \sum_{j} \alpha_j f_j \\
  \text{s.t.}\quad &
    0 \le f_j \le f_{j,\max}, \quad \forall j, \\
  & \sum_{j} f_j \le K_{\text{avg}},
\end{alignat}
\label{LP_problem2}
\end{subequations}
\hspace{-1.7mm}where $K_{\text{avg}}$ bounds the average number of active devices per \gls{tti}. This \gls{lp} is a continuous knapsack problem and admits an intuitive greedy solution as optimal in which devices are ordered by $\alpha_j$ and activated up to their \gls{eh}-limited duty cycle $f_{j,\max}$ until the budget $K_{\text{avg}}$ is exhausted~\cite{boes2023analyzing}. This resulting deployment and duty-cycle allocation provide a convex/heuristic-based method for comparison against the learning-based policy and grid-based heuristics.

\subsection{Hybrid \gls{lp} + \gls{rl} Proposal}\label{sec:LP+RL}
Herein, we use the \gls{lp} solution as an interpretable starting point and then leverage \gls{rl} to refine the solution. 
Note that the entire optimization is performed offline before the devices are physically deployed. 
Specifically, the \gls{lp} stage first determines a static deployment by selecting sensor locations and orientations that maximize coverage over a discretized region of interest and selects the starting duty-cycle fractions for the selected devices while respecting $C_{\text{budget}}$. The outputs from the \gls{lp} stage are then utilized to initialize the \gls{rl} agent rather than starting from a regular grid. Then, once initialized, the \gls{bs} functions as a centralized learning agent and refines the deployment and sensing schedule over multiple simulated episodes, as described in Section~\ref{RL_method}. Unlike the \gls{lp} solution, this method can adapt the positions of the devices, their directional sensing configurations, and individual activity decisions to enhance long-term effective coverage despite \gls{eh} uncertainties. After training, the resulting device locations are fixed, while the learned sensing and duty-cycling policy is applied during operation.

This hybrid approach effectively combines two complementary properties. The \gls{lp} stage offers a structured and interpretable initial operating point, while the \gls{rl} stage captures the nonlinear interactions among coverage overlap, directional sensing, and time-varying energy availability. This method is particularly advantageous in large search spaces because it alleviates the burden on \gls{rl} during the early stages of training compared to purely random or grid-based initialization~\cite{he2022collaborative,zhu2018optimal}. 

\subsection{Complexity Analysis}\label{sec:complexity}
Regarding complexity, the grid-based benchmark has almost negligible optimization complexity because the deployment remains fixed and only a pre-defined duty-cycle pattern is applied. Its offline complexity is, at most, linear in the number of devices, $\mathcal{O}(N)$. The other approaches are more computationally demanding. 
Let $P$ the number of discretized grid points in the region of interest, and $M$ the number of candidate location/orientation configurations used to construct the binary coverage matrix $a_{i,n}$. In addition, let $N_{\mathrm{ep}}$ denote the number of training episodes used by the learning-based methods, and let $|\mathcal{A}|$ denote the effective number of actions explored per device state in the \gls{rl} policy. 
The \gls{lp}-based method follows the formulation described in Section~\ref{sec:LP}. First, constructing the binary coverage matrix $a_{i,n}$ over all grid points and candidate configurations scales as $\mathcal{O}(PM)$. 
Then, the greedy maximum-coverage stage iteratively selects up to $N$ sensing configurations, with dominant cost $\mathcal{O}(NPM)$. 
Meanwhile, the duty-cycle allocation requires sorting the selected devices according to their coverage weights $\alpha_j$, which adds $\mathcal{O}(N\log N)$. Hence, the overall practical complexity of the \gls{lp} method is dominated by the greedy coverage stage and scales as $\mathcal{O}(NPM)$.

The \gls{rl} method adds an iterative training stage. At each episode, the \gls{bs} updates the positions, \gls{fov} orientations, and duty-cycling decisions of the devices according to the policy in Section~\ref{RL_method}, while evaluating the resulting coverage over the discretized region. Therefore, the practical training cost scales as
$\mathcal{O}(N_{\mathrm{ep}}\,N),$
showing that the computational cost of \gls{rl} increases with the number of episodes needed to converge.
The hybrid \gls{lp}+\gls{rl} method combines the two stages and is therefore the most computationally intensive in terms of offline processing per episode, which can be expressed as $\mathcal{O}\!\left(NPM + N_{\mathrm{ep}}\,N\right).$
Although the hybrid \gls{lp}+\gls{rl} method adds an \gls{lp}-based initialization stage before learning, this extra cost is compensated by a substantial reduction in the number of episodes required for convergence. As a result, the total time to convergence can be significantly smaller than that of the standalone \gls{rl} method, despite the higher per-episode computational cost.


\section{Results Analysis}\label{sec:results}

\begin{table}[t]
\caption{Representative \gls{aiot} use cases and $f_{j,\max}$ mapping.}
\label{tab:energy_cases}
\centering
\begin{tabular}{p{0.8cm}p{2.8cm}p{1.8cm}p{1.4cm}}
\hline
\textbf{$f_{j,\max}$} & {Use case}  & Average energy & Peak energy \\
\hline
$10^{-3}$ & Pressure powered switch & $<0.1~\mu$W & $<100~\mu$W \\
$10^{-2}$ & Museum guide & $<1~\mu$W & $<100~\mu$W\\
$10^{-1}$ & Smart Agriculture &  $<100~\mu$W & $<1$ mW \\
\hline
\end{tabular}
\end{table}
The \gls{3gpp} disclosed around 30 representative use cases, some of them discussed in Table~\ref{tab:FoV_classes}. In this paper, we focus on three representative use cases selected according to the energy budget the devices must maintain to remain operational. 
We select values of $f_{j,\max}\in\{10^{-3},10^{-2},10^{-1}\}$ as representative operating regimes derived from the \glspl{kpi} ranges reported for \gls{aiot} use cases in~\cite{3gpp_tr22840}. 
These values are summarized in Table~\ref{tab:energy_cases}. Based on the energy-neutrality condition, the maximum feasible duty cycle of device $j$ is given by
\begin{subequations}
\begin{alignat}{4}
f_{j,\max}&=\min\!\left\{1,\frac{\bar E_j^{\mathrm H}}{E_j^{\mathrm{act}}}\right\},\\
&=\min\!\left\{1,\frac{\lambda_j E_{\mathrm H}}{E_{\mathrm{sense}}+E_{\mathrm{Tx}}}\right\}, 
\end{alignat}
\label{fmax_results}
\end{subequations}  
\hspace{-1.8mm}where $\bar E_j^{\mathrm H}$ denotes the average available energy and  $E_j^{\mathrm{act}}$ the active sensing (reporting) energy. 
The second {line} follows directly from \eqref{f_max}, with $\bar E_j^{\mathrm H}=\lambda_j^{\mathrm H}E_{\mathrm H}$ and $E_j^{\mathrm{act}}=E_{\mathrm{sense}}+E_{\mathrm{Tx}}$. 


We consider an illustrative \gls{aiot} deployment with $N$ devices and evaluate its performance over 150 independent Monte Carlo runs. Each device performs \gls{eh} with $\lambda_j^{\mathrm H}=0.1$. We consider \gls{fov} values of $\{60^\circ,180^\circ,360^\circ\}$, a maximum sensing range of $r_{\max}=3$~m, and an event duration of $t_{\mathrm{ev}}=1$~s.
Table~\ref{table_sim} summarizes the parameters used in simulations unless explicitly stated otherwise. Moreover, we assume the energy buffers are empty at the onset of network deployment and an area resolution of 1~cm. 

All experiments were conducted on a Lenovo ThinkPad laptop equipped with an Intel\textsuperscript{\textregistered} Core\texttrademark{} i7-12700H processor, 16~GB of RAM, and running MATLAB\textsuperscript{\textregistered} R2023a on Windows~11. The large-scale Monte Carlo sweep across all considered scenarios was performed on the University of Oulu's Lehmus high-performance computing environment~\cite{lehmus}, using Nvidia\textsuperscript{\textregistered} V100 GPUs with 16~GB of memory. 
\begin{table}[t!]
\caption{Simulation parameters}
\label{table_sim}
\begingroup \centering
    \begin{tabular}{lll}
    \hline
    \textbf{Parameter}                       & \textbf{Value}     &\textbf{Ref.}\\
    \hline
    $\eta$                  & 1                     & \cite{thomsen2017traffic, ruiz2022energy}\\
    $\lambda^H$  & {0.1}      & \cite{lopez2023energy}\\
    $\xi$                   & $20\times20$          & \cite{thomsen2017traffic,ruiz2022energy}\\
    $f_{j,\max}$            & $\{10^{-3}, 10^{-2}, 10^{-1}\}$    & \cite{3gppTS38391}\\
    \gls{fov}               & $\{60, 180, 360\}$    & \cite{sirmacek2020occupancy,wakai2021deep,ruiz2026context}\\
    $N$                     & $[10^{-2}, 1]$ per $m^2$             & \cite{thomsen2017traffic,ruiz2022energy}\\ 
    $p(d_{i,j})$            & $e^{-\eta d_{i,j}}$   & \cite{thomsen2017traffic, ruiz2022energy}\\
    $r_{\max}$               & 3 meters              & \cite{ul2022learning}\\
    $t_{ev}$                & 1 s
    & \\
    \gls{tti}               & 1 ms              & \\
    \hline
    \end{tabular}{}\\
    \endgroup
\vspace{-2ex}
\end{table}

\subsection{Coverage Performance}
\begin{figure}[t!]
\centering
\includegraphics[width=\linewidth]{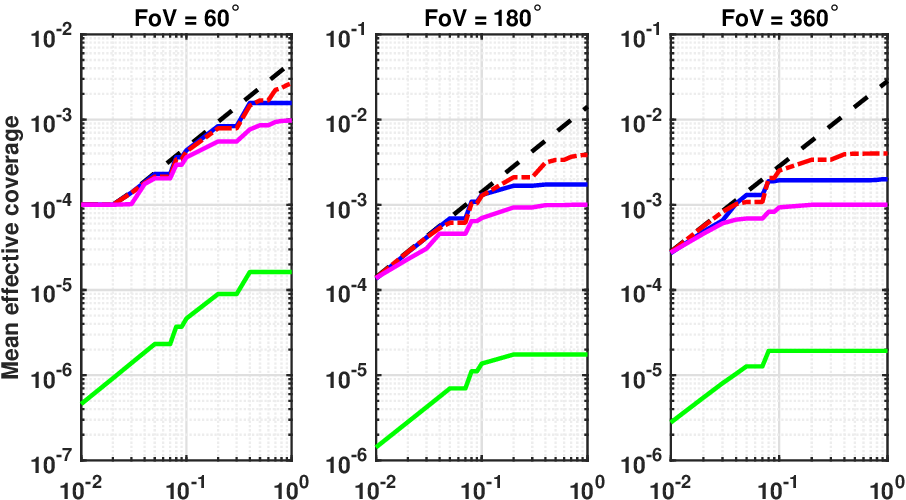}\vspace{3mm}
\includegraphics[width=\linewidth]{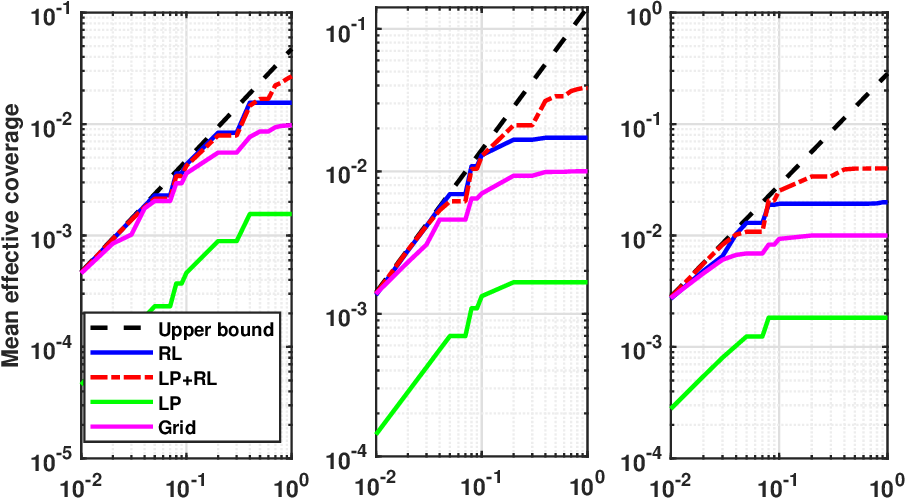}\vspace{3mm}
\includegraphics[width=\linewidth]{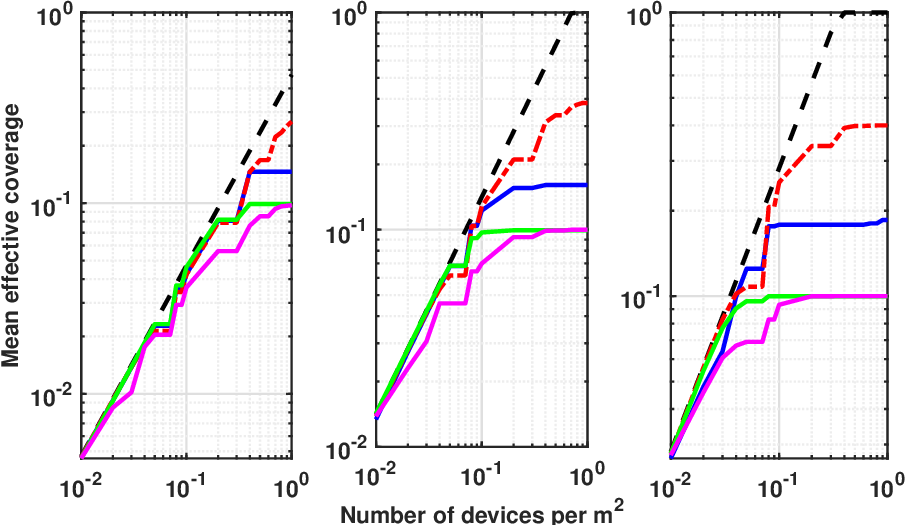}
\caption{Mean effective coverage of the area per \gls{tti} as a function of the device density for different \gls{fov} and duty-cycle regimes. Each column corresponds to a device \gls{fov} configuration, namely $60^\circ$, $180^\circ$, and $360^\circ$, while each row corresponds to a maximum energy-neutral duty cycle $f_{j,\max}\in\{10^{-3},10^{-2},10^{-1}\}$. 
}
\label{fig_results1}
\end{figure}

We consider the grid deployment as a benchmark and assume that each device follows a duty-cycle pattern with active interval $t_{\mathrm{on}}$ and sleep interval $t_{\mathrm{off}}$, chosen so that the average energy consumption does not exceed the harvested energy budget. Accordingly, each device operates under a maximum feasible duty cycle $f_{j,\max}$. 
We also consider an upper bound that assumes the devices are positioned and oriented such that their effective sensing regions cover distinct portions of the region, without unnecessary overlap or boundary losses.  Under these assumptions, the expected covered area is given by the sum of the duty-cycle-weighted effective sensing areas of all devices, capped by the total area of the region of interest. 
For each policy, we evaluate the mean effective coverage per TTI and determine the minimum device density required to achieve different target coverage levels, ranging from low effective coverage to near-complete area coverage.   


\begin{figure*}[t!]
\centering
\includegraphics[width=\linewidth]{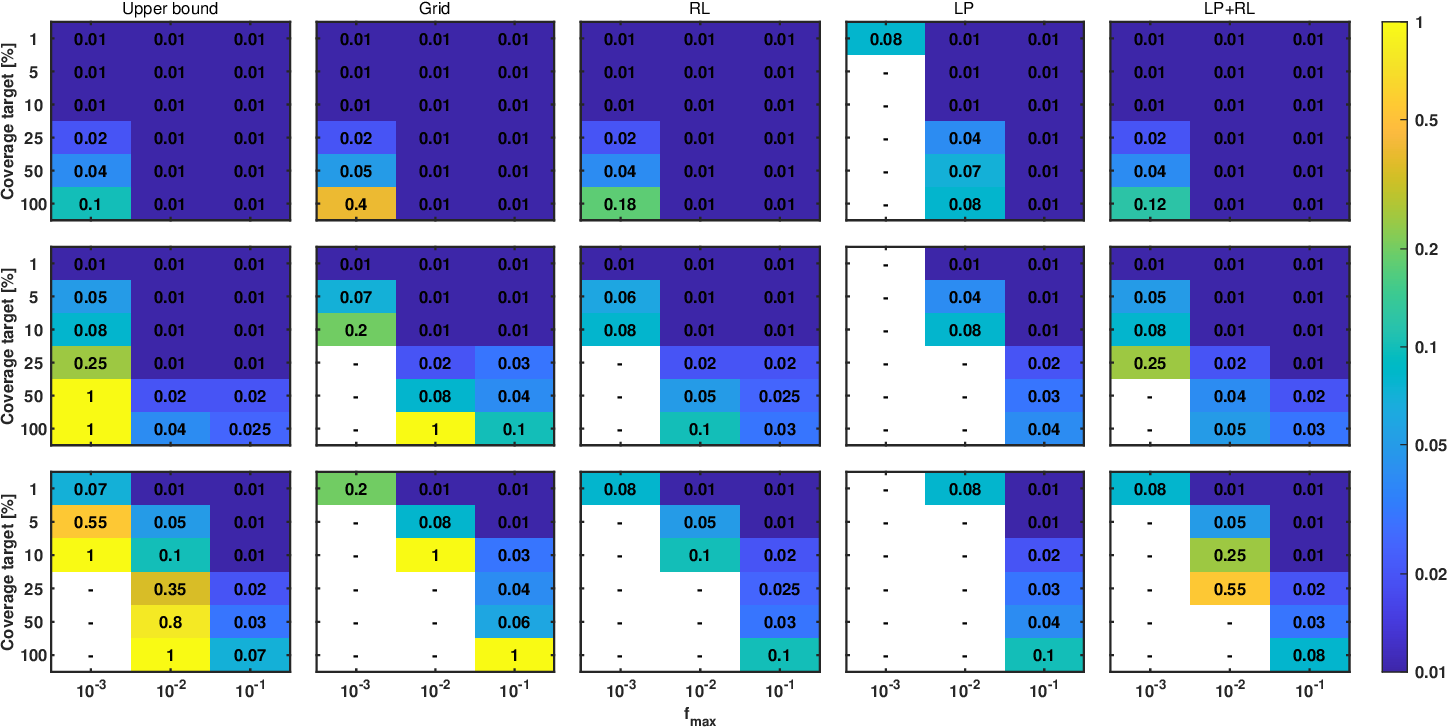}
\caption{Minimum device density (per $m^2$) required to achieve different effective coverage targets for \gls{fov} $180^\circ$ and maximum feasible duty-cycle, $f_{\max}\in\{10^{-3},10^{-2},10^{-1}\}$. Each row corresponds to $t_{ev} \in \{1~\mathrm{s},100~\mathrm{ms},10~\mathrm{ms}\}$ from top to bottom. 
}
\label{fig:min_density}
\end{figure*}

{\figurename~\ref{fig_results1} compares the mean effective coverage achieved by the grid-based benchmark, the \gls{rl}-based policy, the \gls{lp}-based method, and the proposed \gls{lp}+\gls{rl} hybrid against the duty-cycle-aware upper bound, for the different \glspl{fov} and duty-cycle regimes. 
For the most stringent regime, $f_{\max}=10^{-3}$, the \gls{lp} baseline yields the lowest effective coverage despite the \gls{fov}. In contrast, the \gls{lp}+\gls{rl} hybrid consistently achieves the best performance, especially as the number of devices increases. Meanwhile, the  \gls{rl} and grid-based solutions remain in between, with the \gls{rl} policy outperforming the grid-based benchmark.}

{As the duty-cycle increases to $f_{\max}=10^{-2}$ and $10^{-1}$, the performance of the methods separates/distinguishes further. The \gls{lp}+\gls{rl} hybrid is the best-performing scheme for all \glspl{fov}, followed by the  \gls{rl} policy and the grid benchmark, whereas the \gls{lp} solution remains substantially below the others. 
The hybrid \gls{lp}+\gls{rl} solution remains closest to the upper bound, the \gls{rl} method is the second-best solution, and the \gls{lp} method, although improved, still does not match the performance of the hybrid.} 

{Additionally, \figurename~\ref{fig:min_density} quantifies the trends observed in \figurename~\ref{fig_results1} by reporting the minimum device density required to achieve different mean effective-coverage targets for a $180^\circ$ \gls{fov}, across the three duty-cycle regimes and event windows $t_{ev} \in \{1~\mathrm{s},100~\mathrm{ms},10~\mathrm{ms}\}$. The figure shows that extending the event window substantially relaxes the coverage requirements, especially for the medium and high duty-cycle regimes. In particular, for $t_{ev}=1$~s and for $f_{\max}=10^{-2}$ and $f_{\max}=10^{-1}$, most methods already achieve coverage targets up to $10\%$ at the minimum considered density, so the differences among the schemes remain small for low and moderate targets. 
 
Moreover, in the most stringent case, $f_{\max}=10^{-3}$, the  schemes are still clearly separated. In this case, the proposed \gls{lp}+\gls{rl} hybrid and the standalone \gls{rl} policy require the lowest densities to achieve moderate and high coverage targets, whereas the grid benchmark requires noticeably larger densities. The \gls{lp} solution remains the weakest method, as it fails to reach several of the higher target levels within the considered density range. 
Notably, for $f_{\max}=10^{-3}$ and $t_{ev}=10$~ms, the schemes remain limited to roughly 1\%–5\% effective coverage, while the upper bound can reach about 10\%. This highlights the difficulty of guaranteeing reliable sensing under simultaneously stringent energy and latency constraints. In this regime, the combination of a very small feasible duty cycle and a short observation window leaves little opportunity for collaborative sensing, so even optimized practical policies remain far from the theoretical limit. 
As the duty-cycle budget increases to $f_{\max}=10^{-2}$ and $f_{\max}=10^{-1}$, the required density rapidly decreases, showing that most methods, except the standalone \gls{lp}, approach saturation at relatively low densities. 
Overall, \figurename~\ref{fig:min_density} confirms that the proposed \gls{lp}+\gls{rl} strategy remains the most robust and effective solution across the considered duty-cycle regimes and \gls{fov} settings. 
The advantage of the proposed \gls{lp}+\gls{rl} method is especially pronounced for wider \glspl{fov} and moderate duty-cycle budgets, while the \gls{lp} method remains limited by its conservative static duty allocation under stringent energy constraints.

\subsection{Temporal dynamics}
\begin{figure*}[t!]
\includegraphics[width=0.82\linewidth]{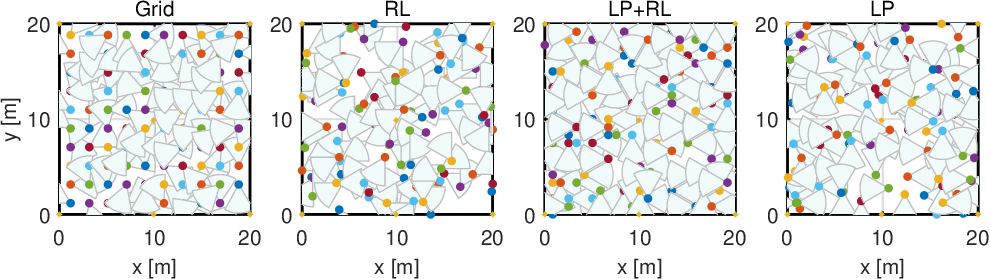}\vspace{2mm}\\
\centering
\includegraphics[width=0.8\linewidth]{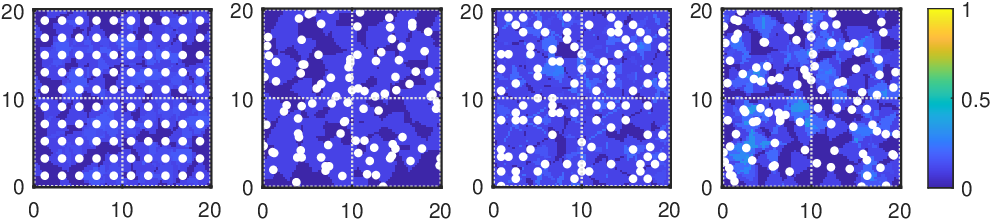}\vspace{2mm}
\includegraphics[width=0.8\linewidth]{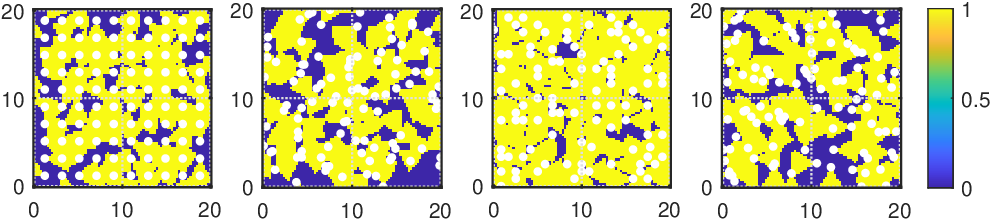}
\caption{Device deployment layouts (top) for the four methods, and spatial time-coverage maps (center and bottom) for a representative scenario with $N=100$ and $\mathrm{FoV}=60^\circ$. The top row shows the final deployments obtained by each method. The middle and bottom rows show the probability of each point being monitored during the observation window for $f_{\max}=0.01$ (center) and $f_{\max}=0.1$ (bottom). 
}
\label{fig_temp1}
\end{figure*}

\begin{figure}[t!]
\centering
\includegraphics[width=0.98\linewidth]{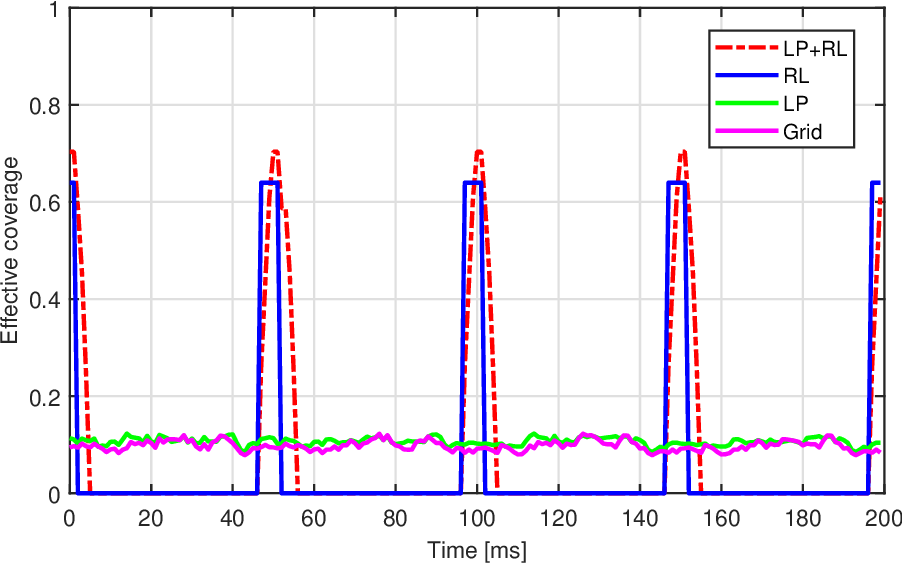}
\caption{Instantaneous effective coverage over a 200 ms observation window for a representative scenario with $N=100$, $\mathrm{FoV}=60^\circ$, and $f_{\max}=10^{-1}$.
}
\label{fig_temp2}
\end{figure}

In this section, we illustrate the instantaneous coverage probability within a time window for a representative scenario with $N=100$ and $\mathrm{FoV}=60^\circ$. Then, we examine the temporal and spatial behavior of the different deployment policies. \figurename~\ref{fig_temp1} shows the device layouts together with heatmaps indicating the probability that the area is covered during $t_{ev}$ for $f_{\max}=\{10^{-2},\ 10^{-1}\}$. \figurename~\ref{fig_temp2} then reports the instantaneous effective coverage over a representative time window of 200~ms for $f_{\max}=10^{-1}$. 
The figures show that the grid-based benchmark preserves a regular and spatially uniform deployment pattern, which provides a predictable sensing structure but also leads to periodic uncovered regions that are visible in the corresponding heatmap. Since the devices follow a static schedule and remain at their initial positions, the resulting time-coverage map exhibits a regular pattern with recurring low-coverage points. 

The \gls{lp}-based solution, although optimized in terms of static placement and duty-cycle allocation, still produces several localized uncovered regions. This highlights the limitations of the \gls{lp}-based approach in sustaining uniform temporal coverage over the entire area. 
The \gls{rl}-based solution enables stronger temporal peaks in effective coverage, as shown in \figurename~\ref{fig_temp2}. However, the corresponding heatmap still reveals uncovered areas, particularly near some border regions and around local spatial gaps introduced by the learned relocation policy. Furthermore, the \gls{lp}+\gls{rl} method provides the best temporal behavior. The hybrid strategy combines the structured initialization of the \gls{lp} stage with the adaptive refinement of the \gls{rl}, leading to a more balanced configuration. This is reflected in the heatmap of \figurename~\ref{fig_temp1}, where the covered area is more uniformly distributed, and the number of persistent blind regions is reduced compared with the other methods. Notably, in \figurename~\ref{fig_temp2}, the \gls{lp}+\gls{rl} method also attains the highest instantaneous effective coverage, followed by the standalone \gls{rl} policy, whereas the \gls{lp} and grid baselines remain at much lower levels throughout the observation window. 

Note that, in \figurename~\ref{fig_temp2}, the \gls{rl} and \gls{lp}+\gls{rl} methods generate periodic peaks in effective coverage, which correspond to coordinated sensing opportunities when a larger fraction of devices becomes active. Between these peaks, the effective coverage drops to lower values, reflecting the sleep phases required to maintain sustainable operation. In contrast, the grid and \gls{lp} baselines exhibit flatter temporal trajectories with considerably smaller peaks, indicating that their static duty-cycle design is less effective at exploiting the available energy budget to create high-coverage sensing opportunities.

\subsection{Sensitivity Analysis}
The \gls{rl} approach depends on the reward function to balance coverage and energy efficiency. Increasing the weighting factor $\mu$ prioritizes maximizing coverage, but simultaneously reduces the relative importance of the term that encourages operation within the available energy budget. As a result, a large $\mu$ can improve short-term coverage while compromising long-term energy sustainability. Therefore, selecting an appropriate value of $\mu$ is essential to achieve a suitable trade-off between instantaneous coverage performance and sustained covered area over time. 

\figurename~\ref{fig_sens_cov} illustrates the sensitivity of the achieved coverage to the reward weight, $\mu$, for both the  \gls{rl} method and the proposed \gls{lp}+\gls{rl} hybrid, for $N=100$, $\mathrm{FoV}=60^\circ$, and $f_{\max}=0.1$. The results show that the hybrid approach consistently attains higher coverage than the  \gls{rl} method over the entire range of $\mu$ values considered. In particular, the \gls{lp}+\gls{rl} method's coverage remains concentrated around 0.69–0.71, whereas the \gls{rl} policy remains between $0.62$ and $0.64$. This indicates that the structured \gls{lp} initialization provides a favorable operating point that is already well aligned with the spatial coverage objective, allowing the subsequent learning stage to refine the solution without strong dependence on the exact reward tuning. 

Moreover, the \gls{rl} method appears more sensitive to the choice of $\mu$. As $\mu$ increases, its coverage tends to decrease while the error bars become larger. Therefore, without a strong initialization, emphasizing the coverage term alone does not necessarily improve the final deployment. Instead, it can weaken the contribution of the energy-related component and lead toward less balanced deployment and scheduling decisions. In contrast, the \gls{lp}+\gls{rl} hybrid shows only mild variation across different values of $\mu$, highlighting its robustness to reward-weight selection, indicating that most of the achievable coverage is already captured by the \gls{lp}-based initialization. Notably, the hybrid method not only improves average performance but also reduces sensitivity to hyperparameter tuning, which is desirable for scalable optimization in large \gls{aiot} deployments. 

\begin{figure}[t!]
\centering
\includegraphics[width=0.98\linewidth]{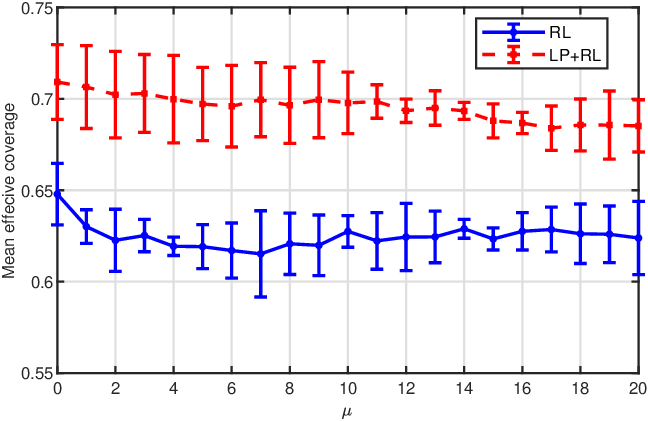}
\caption{Mean effective coverage as a function of the reward weight $\mu$ for the \gls{rl} and \gls{lp}+\gls{rl} methods, for $N=100$, $\mathrm{FoV}=60^\circ$, and $f_{\max}=10^{-1}$. Error bars indicate the standard deviation of the effective coverage across runs.
}
\label{fig_sens_cov}
\end{figure}

\begin{figure}[t!]
\centering
\includegraphics[width=0.98\linewidth]{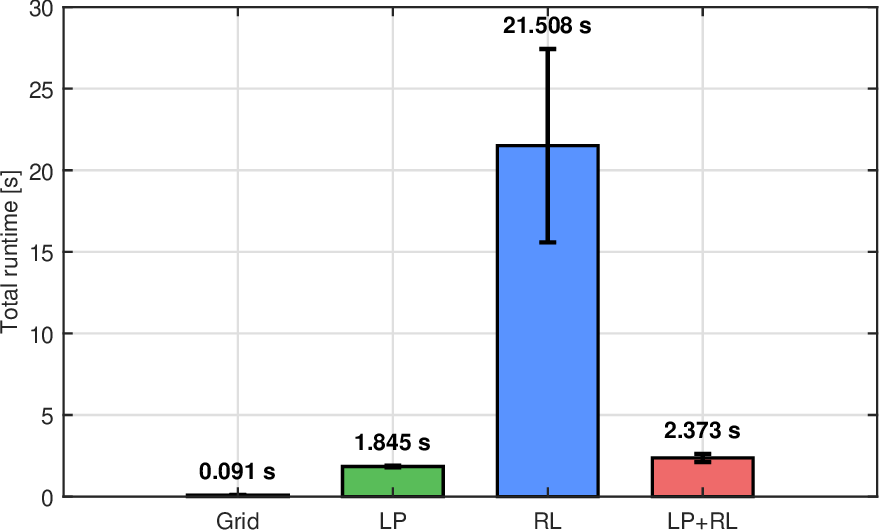}
\caption{Average runtime per scenario for the four methods.
}
\label{fig:complexity}
\end{figure}

\subsection{Convergence Analysis}
In this section, we calculate the total convergence time of each proposed method to support the complexity analysis in Section~\ref{sec:complexity}. In our implementation, the average total runtimes were $t_{\mathrm{Grid}}=0.091$~s, $t_{\mathrm{LP}}=1.845$~s, $t_{\mathrm{RL}}=21.508$~s, and $t_{\mathrm{LP+RL}}=2.373$~s, as shown in \figurename~\ref{fig:complexity}. Herein, we assume convergence for \gls{rl} as the first episode at which the smoothed reward remains within $1\%$ of its final reference value for 10 consecutive episodes. Specifically, convergence occurs when
\begin{equation}
\Delta r_{ep} =
\frac{\left|\bar{r}_{ep}-r_{\mathrm{ref}}\right|}
{\max\left\{|r_{\mathrm{ref}}|,\epsilon\right\}}
\leq 0.01,
\end{equation}
where $\bar{r}_{ep}$ is the smoothed reward at episode $ep$, $r_{\mathrm{ref}}$ is the average smoothed reward over the final training window, and $\epsilon = 10^{-5}$ is used for stability to avoid division by zero. 

These results are consistent with the complexity analysis, showing that the grid and standalone \gls{lp} baselines are the least computationally and time demanding, respectively.  
Moreover, although the hybrid method is the most computationally expensive per episode, it shows convergence much faster than the standalone \gls{rl}. Consequently, it reduces the total offline optimization time by nearly one order of magnitude. As discussed in Section~\ref{sec:complexity}, the structured \gls{lp} initialization not only improves robustness but also accelerates convergence by reducing the exploration required during learning. Note that these runtimes are implementation- and platform-dependent and are therefore reported only as indicative empirical costs of the considered implementations.


\section{Conclusion}\label{sec:conclusions}
We explored the deployment and sensing scheduling of \gls{aiot} devices with directional sensing capabilities subject to strict energy constraints. We developed a sensing framework that integrates distance-dependent detection probabilities with an angular \gls{fov}, and we formulated a long-term coverage maximization problem under energy-neutral conditions. We compared four different solution strategies: a grid deployment with static duty cycling, a \gls{lp} baseline with an energy-aware duty-cycle design, a \gls{rl} policy that dynamically adjusts device locations and duty cycles starting from a grid layout, and a hybrid \gls{lp}+\gls{rl} method.
Numerical evaluations of representative A-IoT use cases demonstrated that the hybrid \gls{lp}+\gls{rl} policy is the most robust practical solution across the considered duty-cycle regimes and \gls{fov} settings. It consistently achieves the highest mean effective coverage, particularly in low and moderate {\gls{eh}} regimes. The runtime analysis showed that the structured \gls{lp} initialization substantially accelerates learning, reducing the total offline optimization time {by up to 89\% compared to the standalone \gls{rl}.} 
Moreover, the standalone \gls{rl} policy generally performs better than the grid benchmark, while the standalone \gls{lp} method is constrained by its conservative static duty-cycle allocation, particularly under tight energy constraints.
The results indicate that combining optimization-based initialization with learning-based refinement is an effective strategy for energy-constrained directional sensing networks, improving both coverage performance and robustness. 



\bibliographystyle{IEEEtran}
\bibliography{bib}



\end{document}